\newif\ifdouble
\newif\ifsingle
\newif\ifchange

\documentclass[sigconf, usenames, dvipsnames]{acmart}
\doublefalse  

\newcommand{\imgparam}{1.0}

\usepackage{listings}
\usepackage{xcolor} 

\usepackage{dblfloatfix} 
\usepackage{booktabs} 
\usepackage{arydshln}
\usepackage{subfig}
\usepackage{booktabs}
\usepackage{multirow}
\usepackage{makecell}
\newcommand{\remove}[1]{{\color{red}{\sout{#1}}}}

\renewcommand{\remove}[1]{}

\AtBeginDocument{%
  \providecommand\BibTeX{{%
    \normalfont B\kern-0.5em{\scshape i\kern-0.25em b}\kern-0.8em\TeX}}}

\author{Keiichi Ihara}
\affiliation{%
\institution{University of Colorado Boulder}
\city{Boulder}
\state{Colorado}
\country{United States}}
\affiliation{%
\institution{Tohoku University}
\city{Sendai}
\country{Japan}}
\email{keiichi.ihara@colorado.edu}

\author{DaeHo Lee}
\affiliation{%
\institution{Department of AI Convergence, Gwangju Institute of Science and Technology}
\city{Gwangju}
\country{Republic of Korea}}
\email{leedaeho@gm.gist.ac.kr}

\author{Manato Abe}
\affiliation{%
\institution{Tohoku University}
\city{Sendai}
\state{Miyagi}
\country{Japan}}
\email{riec-icd-office@grp.tohoku.ac.jp}

\author{Hye-Young Jo}
\affiliation{%
\institution{University of Colorado Boulder}
\city{Boulder}
\state{Colorado}
\country{United States}}
\email{hye-young.jo@colorado.edu}

\author{Ryo Suzuki}
\affiliation{%
\institution{University of Colorado Boulder}
\city{Boulder}
\state{Colorado}
\country{United States}}
\affiliation{%
\institution{Tohoku University}
\city{Sendai}
\country{Japan}}
\email{ryo.suzuki@colorado.edu}

\begin{document}

\newcommand{\system}{CinemaWorld}


\title{\system{}: Generative Augmented Reality with LLMs and 3D Scene Generation for Movie Augmentation}

\begin{abstract}
We introduce \system{}, a generative augmented reality system that augments the viewer’s physical surroundings with automatically generated mixed reality 3D content extracted from and synchronized with 2D movie scenes. Our system preprocesses films to extract key features using multimodal large language models (LLMs), generates dynamic 3D augmentations with generative AI, and embeds them spatially into the viewer’s physical environment on the Meta Quest 3. To explore the design space of \system{}, we conducted an elicitation study with eight film students, which led us to identify several key augmentation types, including particle effects, surrounding objects, textural overlays, character-driven augmentation, and lighting effects. We evaluated our system through a technical evaluation (N=100 video clips), a user study (N=12), and expert interviews with film creators (N=8). Results indicate that \system{} enhances immersion and enjoyment, suggesting its potential to enrich the film-viewing experience.
\end{abstract}



\begin{teaserfigure}
\centering
\includegraphics[width=\textwidth]{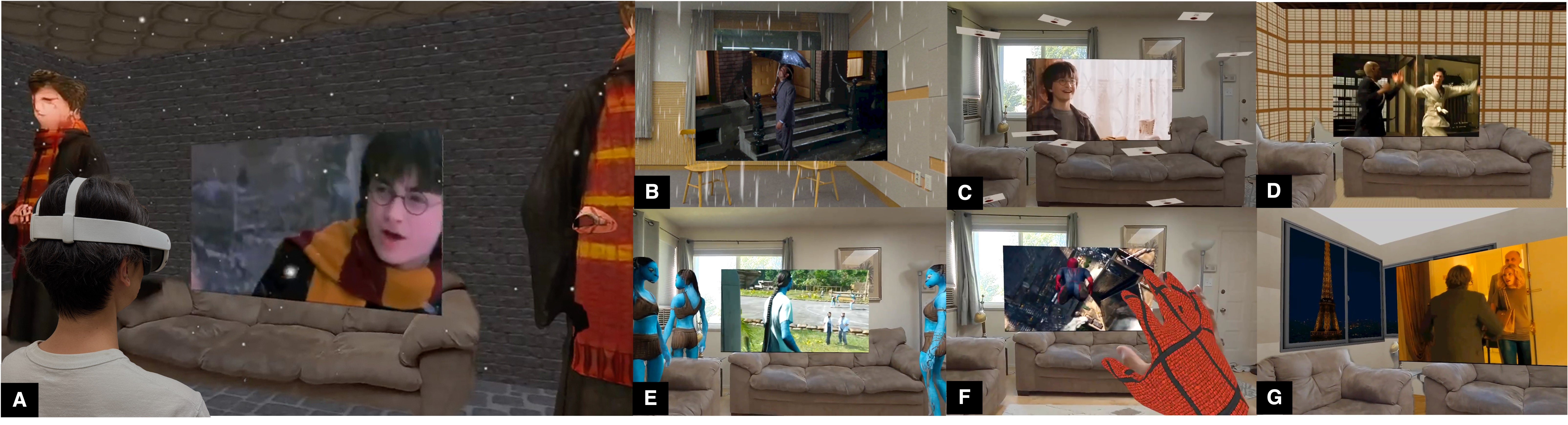}
\caption{\system{} is a generative augmented reality system that augments the viewer's physical environment in synchrony with the film, leveraging generative AI (a). Our system combines several key augmentations, such as particle effects (b), surrounding objects (c), room textures (d), character presence (e), body transformation (f), window augmentation (g), and lighting effects.}
\label{fig:teaser}
\end{teaserfigure}
\maketitle
\section{Introduction}
With the rise of generative AI and large language models (LLMs), it is now possible to create dynamic augmented/mixed reality (AR/MR\footnote{In this paper, we use AR and MR interchangeably, or treat MR as a subset of AR, based on the definitions discussed in \cite{speicher2019mixed}.}) experiences \textit{on demand}, without manual preparation or programming. Historically, mixed reality experiences had to be carefully prepared and predesigned; today, however, we can start \textit{generating} such experiences automatically through AI-generated assets, textures, and programs that seamlessly blend into the real-world environment. We call this emerging paradigm \textit{\textbf{generative augmented reality (GenAR)}}, or more broadly, generative extended reality (GenXR), in which prior work explores parts of this space, including content augmentation~\cite{gunturu2024augmented, chulpongsatorn2023augmented}, VR and AR scene generation~\cite{cheng2019vroamer, sra2016procedurally, lee2025imaginatear, de2024llmr}, and generative augmented instruction~\cite{shi2025caring, zhao2025guided}. 

This paper pushes this boundary further in the context of \textit{movie augmentation}, focusing on how a mixed reality scene can be generated to transform the physical environment based on the current 2D film content. Our system takes a movie scene in MR and \textit{automatically} generates embedded 3D effects by orchestrating multimodal LLMs, texture/object generation, and real-world embedding. As illustrated in Figure~\ref{fig:teaser}, the system dynamically transforms the room with film-synchronized effects, such as particles, objects, room textures, and characters, turning the real world into an extension of the movie world. Unlike previous systems such as \textit{IllumiRoom}~\cite{jones2013illumiroom} and \textit{ExtVision}~\cite{kimura2018extvision}, our system introduces two key novelties: 1) \textit{leveraging generative AI}, which enables automatic analysis and generation for movie augmentation at scale, and 2) \textit{blending 3D content into mixed reality scene}, which provides immersive spatial augmentation by seamlessly integrating generated 3D objects with room elements such as floors, walls, and even the viewer’s body.

This work makes three main contributions. First, we contribute to the design space of \system{} by identifying key augmentation strategies for MR-based cinematic experiences. We conducted a formative elicitation study with eight film students, exploring how physical surroundings could be augmented based on a film's visual elements. From this, we identified seven key augmentation methods: 1) particle effects, 2) surrounding objects, 3) room textures, 4) character presence, 5) body transformation, 6) window augmentation, and 7) lighting effects.

Second, we contribute a novel film-to-MR augmentation pipeline. Building on the identified design space, our system follows a three-stage process. First, it analyzes the movie scene using large vision–language models (e.g., Gemini) to extract a semantic description that includes timestamp, scene context, visual effect types, objects in the scene, background textures, and lighting. Second, it captures the user’s physical environment via mixed reality scanning on the Meta Quest 3 (fast surface detection) or high-fidelity LiDAR scanning (detailed 3D meshes with semantic segmentation). Third, it generates augmentations, including room textures, lighting, animated particles, and 3D objects, using generative AI (ChatGPT-4o image models, InstantMesh for 3D), and renders them on the Meta Quest 3, spatially aligned with the detected room geometry and synchronized with the movie timeline. 

Third, we contribute insights from three evaluations: a technical evaluation, a usability study (N=12), and expert interviews with film creators (N=8). 
In the technical evaluation, we tested our system using 100 video clips, assessing visual, semantic, and temporal alignment across seven augmentation types, resulting in 21 evaluation types. Among these, 18 types achieved an accuracy higher than 80\%, indicating that most features are functioning reliably. For the remaining three conditions with lower accuracy, we analyze the causes, which include errors in video analysis and failures in generating intended images.

In the usability study, we qualitatively compared our approach against both a no-augmentation baseline and a 2D augmentation method (e.g., \textit{ExtVision}~\cite{kimura2018extvision}) to investigate how our system enhances the film viewing experience. Results showed that our system increased immersion and enjoyment compared to the other two conditions. 

The expert interviews highlight the potential of our system to transform the viewing experience, allowing audiences to feel as if they are stepping onto the stage of the film and turning passive watching into a more engaging experience. For documentaries, it can also enrich understanding by situating viewers within historical or cultural contexts, deepening comprehension of the film’s intent. At the same time, filmmakers raised concerns about control and authorship, particularly how to align augmentations with the creator’s vision while avoiding distraction. Based on their feedback, we discuss future directions such as providing authoring tools for creators and mechanisms for adjusting the augmentation effects.

In summary, this paper makes the following contributions:
\begin{itemize}
  \item A design space of \system{}, derived from a formative elicitation study with eight film students.
  \item The implementation and technical details of \system{}, a generative system that augments physical environments based on film content by leveraging generative AI.
  \item Findings and insights from a technical evaluation, a user study (N=12), and expert interviews with film creators (N=8).
\end{itemize}

\begin{figure*}[t]
\centering
\includegraphics[width=\textwidth]{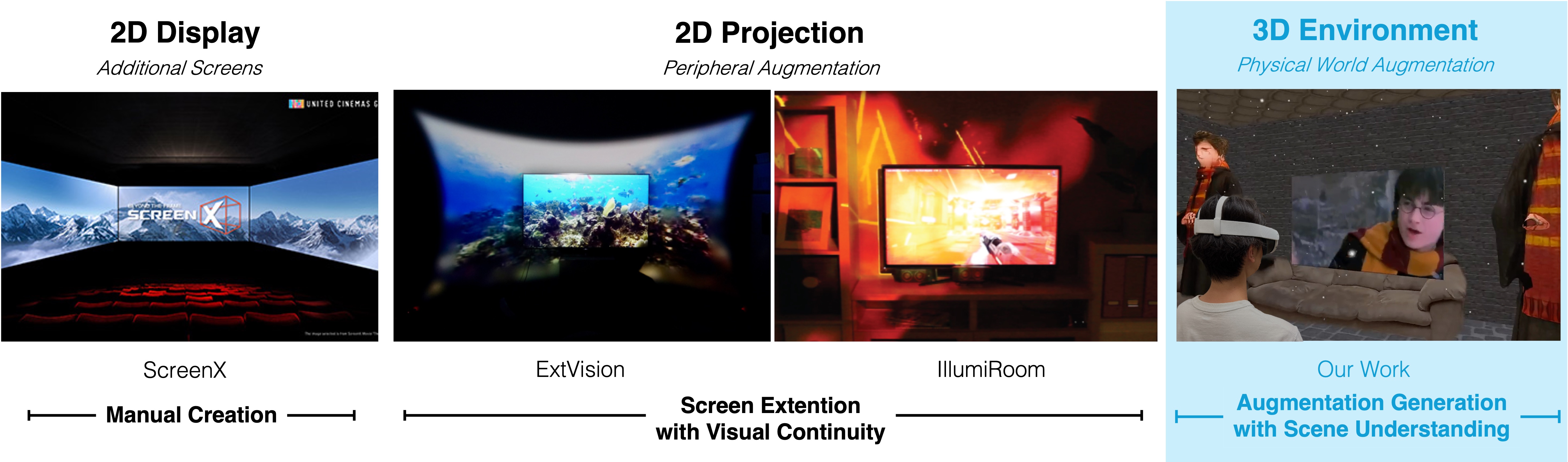}
\caption{
Compared to existing approaches (e.g., \textit{ScreenX}~\cite{ScreenX23:online}, \textit{ExtVision}~\cite{kimura2018extvision}, \textit{IllumiRoom}~\cite{jones2013illumiroom}), our work extracts key elements from movie scenes using scene understanding and generates movie-synchronized 3D mixed reality content in the viewer's physical environment.
}
\label{fig:designspace}
\end{figure*}

\section{Related Work}

\subsection{Augmenting Film and TV Experiences}
A variety of technologies have been developed to enhance the film and television viewing experience, aiming to increase enjoyment and immersion. These approaches range from conventional 2D screen-based systems to projection-based displays and head-mounted displays (HMDs).

2D screen-based augmentations have been introduced primarily in theater settings. For example, \textit{ScreenX}~\cite{ScreenX23:online} extends the field of view by adding additional screens to the left and right of the main screen. In another approach, \textit{COSM}~\cite{Cosm} and \textit{SPHERE}~\cite{Sphere} utilize half-dome screens to achieve a wide field-of-view, surrounding the viewer with visuals that enhance immersion. In addition, \textit{4DX}~\cite{4DX} enhances immersion through multisensory feedback such as motion, scent, and vibration.

Projection-based systems have also been explored as an alternative means of augmenting the viewing experience~\cite{jones2013illumiroom, kimura2018extvision, vatavu2012point, leonidis2021going, vatavu2013there}. \textit{IllumiRoom}~\cite{jones2013illumiroom} and \textit{ExtVision}~\cite{kimura2018extvision}, for example, extended visual content into the peripheral area around the TV via projection. \textit{Going Beyond Second Screens}~\cite{leonidis2021going} projected related content onto surrounding physical surfaces, such as tables and walls, while \textit{Around-TV}~\cite{vatavu2013there} displayed supplementary widgets and web content within the space near the television.

Head-mounted displays (HMDs), such as VR and MR headsets, have been utilized to enhance film and television viewing experiences~\cite{baillard2017multi, vinayagamoorthy2019personalising, saeghe2019augmenting}. For instance, \textit{Multi Device Mixed Reality TV}~\cite{baillard2017multi} introduced applications including 3D models of TV-related products, virtual subtitles, and a contextual TV guide displayed around the screen. Similarly, Vinayagamoorthy et al.~\cite{vinayagamoorthy2019personalising} presented a system that overlays virtual sign language interpreters within the AR space surrounding the television frame.

Building on previous work, our work introduces two key novelties: 1) leveraging generative AI to analyze movie content and automatically generate contextual augmentations, and 2) integrating 3D movie augmentations into the viewer’s 3D physical environment using mixed reality. This approach enables scalable generation of film-specific augmentations while enhancing spatial immersion through 3D integration.

\subsection{Blending Real and Virtual Worlds}
Prior work has explored various methods of blending the physical and virtual worlds, including effectively placing virtual content in real spaces, bringing physical elements into virtual environments, and transforming the surrounding space into playful, immersive experiences.

Researchers have explored how to effectively utilize the physical environment when placing virtual content. They proposed several approaches to anchor virtual objects, ranging from leveraging simple physical edges and surfaces (e.g., \textit{SnapToReality}~\cite{nuernberger2016snaptoreality}, Ens et al.'s work~\cite{ens2015spatial}) to using complex geometric alignment (e.g., \textit{HeatSpace}~\cite{fender2017heatspace}, \textit{OptiSpace}~\cite{fender2018optispace}, \textit{SpaceState}~\cite{fender2019spacestate}). Other methods leverage object-level integration (e.g., \textit{BlendMR}~\cite{han2023blendmr}) and semantic positioning (e.g., \textit{SemanticAdapt}~\cite{cheng2021semanticadapt}, \textit{SituationAdapt}~\cite{li2024situationadapt}).

On the other hand, several works have explored bringing physical objects into virtual environments to enable users to interact with physical objects and nearby people while immersed in VR. These studies have integrated various physical elements such as keyboards~\cite{knierim2018physical}, passersby~\cite{von2019you}, and everyday items such as cups~\cite{budhiraja2015s}.

Moreover, researchers have explored how to transform the physical environment into an immersive and playful space. 
Researchers used projection mapping to bring 2D content beyond the screen and onto physical surfaces, effectively expanding the screen into the physical environment (e.g., \textit{IllumiRoom}~\cite{jones2013illumiroom}, \textit{RoomAlive}~\cite{jones2014roomalive}, \textit{Room2Room}~\cite{pejsa2016room2room}, \textit{Dyadic Projected SAR}~\cite{benko2014dyadic}, and \textit{OptiSpace}~\cite{fender2018optispace}). Researchers have also explored how to virtually interact with physical objects in the environment using VR/MR headsets, focusing on modifying (e.g., \textit{SceneCtrl}~\cite{yue2017scenectrl} and \textit{Diminished Reality}~\cite{cheng2022towards, mori2017survey}) and replacing them (e.g., \textit{Substitutional Reality}~\cite{simeone2015substitutional, suzuki2012substitutional}, \textit{SAVE}~\cite{hoster2023style}, \textit{RealityCheck}~\cite{hartmann2019realitycheck}, \textit{TransforMR}~\cite{kari2021transformr}, \textit{DreamWalker}~\cite{yang2019dreamwalker}, \textit{VRoamer}~\cite{cheng2019vroamer}, and Sra et al.'s work~\cite{sra2016procedurally}).
Furthermore, researchers have explored interacting with the entire physical environment through more complex forms of interaction, such as temporal manipulation and on-demand visualization (e.g., \textit{Mixed Voxel Reality}~\cite{regenbrecht2017mixed}, \textit{Remixed Reality}~\cite{lindlbauer2018remixed}, \textit{Virtual Reality Annotator}~\cite{ribeiro2018virtual}, and \textit{RealityEffects}~\cite{liao2024realityeffects}).

In this work, we blend movie content with the physical environment by leveraging generative AI. The environment is transformed into a playful cinematic world through augmentations that are automatically generated based on key elements of the film, extracted using a scene understanding AI.

\subsection{Generative AI for Film Creation}
Generative AI (GenAI) has recently been adopted across various stages of film production, including scriptwriting, storyboard creation, and video generation. Tools such as \textit{Veo}~\cite{VeoGoogl53:online}, \textit{Sora}~\cite{SoraOpen45:online}, and \textit{Runway}~\cite{RunwayTo17:online} enable direct video generation from scripts. Similarly, \textit{LTX Studio}~\cite{ltxstudio} and \textit{Katalist}~\cite{katalist} support automated storyboard generation, streamlining pre-production workflows. More specialized tools assist in post-production. For instance, \textit{ILM FaceSwap}~\cite{ilm} allows for face replacement, \textit{Flawless AI}~\cite{FlawlessAI} aligns lip movements with dubbed dialogue, and \textit{Flow Studio}~\cite{FlowStudio} enables integration of 3D CGI characters into 2D footage. For writing tasks, general-purpose tools such as \textit{ChatGPT}~\cite{ChatGPT} and \textit{Gemini}~\cite{Gemini} are widely used, while \textit{NovelAI}~\cite{NovelAI} and \textit{AI Dungeon}~\cite{AIDungeon}  are more tailored to storytelling-oriented writing.

In HCI research, researchers have explored how GenAI can support video creation workflows. For instance, \textit{LAVE}~\cite{wang2024lave} leverages an LLM-powered agent that plans and executes based on the user’s objectives. \textit{Lotus}~\cite{barua2025lotus} supports the creation of short-form videos from long-form videos by using summarization generated from LLMs. \textit{MovieFactory}~\cite{zhu2023moviefactory} leverages LLM to expand textual prompts into detailed scripts, from which it generates corresponding audiovisual content.

Futhermore, researchers have recently begun to investigate the sociological aspects of these tools, focusing on how creators use them, what attitudes they hold, and what challenges arise~\cite{tang2025understanding, kim2024unlocking, halperin2025underground}. For example, Tang et al.~\cite{tang2025understanding} interviewed screenwriters to understand current practices and perceptions of generative AI in screenwriting. Kim et al.~\cite{kim2024unlocking} examined the use of generative AI among creators of short-form videos, while Halperin et al.~\cite{halperin2025underground} studied amateur filmmakers and their engagement with these technologies.

However, prior work has primarily examined generative AI during the production process. Less attention has been given to its potential for enhancing the viewing experience after production. In this work, we leverage generative AI to augment the viewer's physical surroundings in synchronization with movies, by extracting key elements from films and generating mixed reality 3D content.

\subsection{Generative AR}
Researchers have started studying the automatic generation of mixed reality experiences with generative AI, which we call generative augmented reality (GenAR). Given a growing interest in the intersection of AI and AR~\cite{hirzle2023xr, suzuki2023xr, tang2025llm}, researchers have started exploring integrating generative AI into AR interfaces. For instance, researchers have explored a variety of approaches in VR and AR scene generation, such as creating 3D props (e.g., \textit{LLMR}~\cite{de2024llmr}, \textit{GesPrompt}~\cite{hu2025gesprompt}, \textit{LLMER}~\cite{chen2025llmer}), designing 3D layouts (e.g., \textit{VRCopilot}~\cite{zhang2024vrcopilot}), generating outdoor environments (e.g., \textit{ImaginateAR}~\cite{lee2025imaginatear}, \textit{VRoamer}~\cite{cheng2019vroamer}, Sra et al.’s work~\cite{sra2016procedurally}), and constructing VR scenes (e.g., \textit{DreamCodeVR}~\cite{giunchi2024dreamcodevr}). For content augmentation, researchers have also investigated how to add information and interactions to existing media, such as augmenting textbooks (e.g., \textit{Augmented Physics}~\cite{gunturu2024augmented}, \textit{Augmented Math}~\cite{chulpongsatorn2023augmented}, \textit{RealitySummary}~\cite{gunturu2024realitysummary}) and presentation (e.g., \textit{RealityTalk}~\cite{liao2022realitytalk}). Regarding interaction with physical objects, prior work has proposed obtaining generative outputs through interactions with physical objects (e.g., \textit{GazePointAR}~\cite{lee2024gazepointar}, \textit{AiGet}~\cite{cai2025aiget}) and generating user interfaces for physical objects (e.g., \textit{XR-Objects}~\cite{dogan2024augmented}). In the context of generating instructions, researchers have also developed methods to automatically convert text descriptions or videos into engaging 3D instructions (e.g., \textit{Guided Reality}~\cite{zhao2025guided}, \textit{CARING-AI}~\cite{shi2025caring}, \textit{Video2MR}~\cite{ihara2025video2mr}).

Our work builds upon these prior studies by extending this line of research to the domain of movie augmentation. We propose a system that generates a film-synchronized, mixed reality enhanced physical environment from movies. Specifically, we leverage multi-modal LLMs to extract key elements from films, image and object generation models to create 3D props for augmentation, and 3D capturing and segmentation technologies to analyze the surrounding physical environment.

\section{Design Elicitation}
To explore the design space of mixed reality movie augmentation, we conducted a formative design elicitation study with eight film students.
The goal of this study was to elicit examples of how the physical environment could be modified and visualized in response to film content.

\subsection{Method}
\subsubsection{Participants}
We recruited eight participants (5 males, 3 females, aged 19–27, M = 21.3, SD = 2.49) from two film schools. Four participants were third-year undergraduates, and the other four were second-year undergraduates. Each study session lasted approximately two hours and was conducted via individual Zoom meetings.  Participants received a \$20 Amazon gift card as compensation for their participation.


\subsubsection{Protocol}
First, we explained the goals of the study and introduced the idea of mixed reality movie augmentation. Participants were then asked to show film clips they would be interested in augmenting. We watched the selected clips together, and participants were encouraged to pause the video whenever they had an idea for how the physical environment could be augmented in response to the content. To visualize the proposed augmentations, we collaboratively generated visual mock-ups. Participants described the scene and the augmentation idea in natural language, which we used as prompts for ChatGPT 4o Image Generation. A reference image depicting a relevant setting (e.g., room, train, car) was included to provide contextual grounding. The resulting images were imported into PowerPoint for post-editing. This step allowed us to adjust visual details such as brightness, color balance, and spatial placement of virtual elements, thereby enhancing the fidelity and clarity of the mock-ups.

\begin{figure*}[t]
\centering
\includegraphics[width=\textwidth]{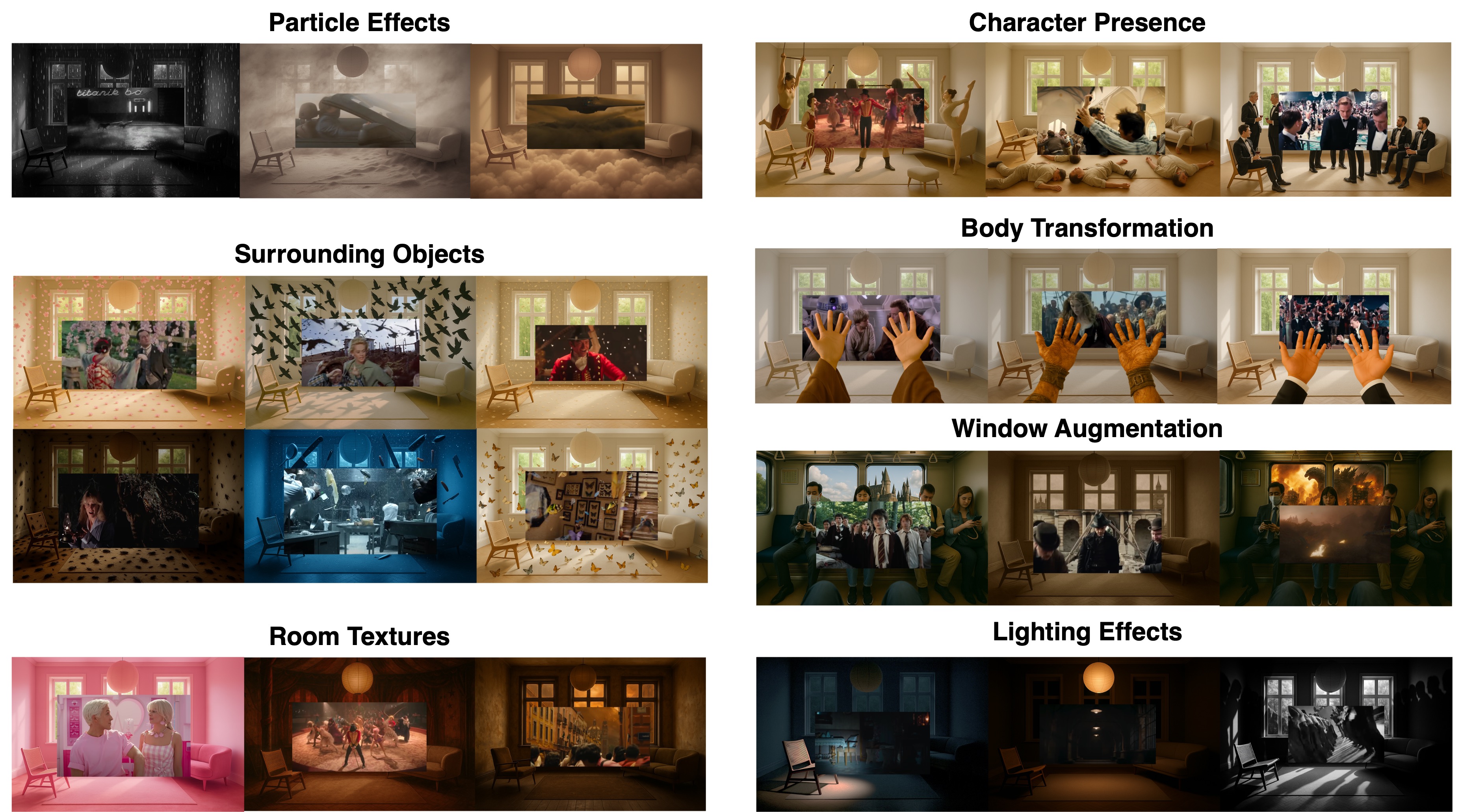}
\caption{
Design space of \system{}. Example images were collaboratively created with participants.}
\label{fig:designspace}
\end{figure*}
\subsection{Elicited Augmentation Ideas}
We collected a total of 43 design suggestions from participants regarding how the physical environment could be augmented in response to specific film scenes.
Three authors reviewed the suggestions and employed an affinity diagramming approach to cluster similar ideas.
This process resulted in the identification of key dimensions that characterize the design space of \system{}, which are detailed in the following section.
Figure~\ref{fig:designspace} illustrates this design space along with example images that were collaboratively created with participants during the study.


\subsubsection*{\textbf{Particle Effects (7/43)}}
Participants expressed interest in overlaying virtual particles onto the physical environment to convey the atmosphere of specific cinematic scenes. For instance, in stormy weather scenes, rendering wind-blown rain within the room was considered compelling, as it could increase immersion. In scenes involving blowing sand, participants proposed simulating airborne particles that obscure visibility, thereby enhancing realism. Some participants also mentioned scenes set high above the clouds. In such cases, placing clouds along the bottom of the room could evoke the sensation of floating above them, effectively reinforcing the spatial setting. 

\subsubsection*{\textbf{Surrounding Objects (11/43)}}
Participants highlighted the value of augmenting scenes with scene-relevant objects as a means to enhance immersion. In contrast to atmospheric particles, this form of augmentation involves larger elements, such as animals and everyday objects, rather than subtle environmental cues like dust or rain.
These objects can help establish both the spatial and temporal context of a scene. For example, falling cherry blossoms may suggest that the scene takes place in Japan during spring, while floating confetti may indicate a festive setting, such as a party or parade.
Participants also expressed interest in populating the room with unsettling creatures such as aggressive birds or grotesque insects to convey horror or tension. Additionally, the manipulation of object motion was considered an effective way to communicate altered perceptions of time, such as when the world appears paused or slowed.

\subsubsection*{\textbf{Room Textures (9/43)}}
Participants also emphasized the importance of altering the texture of the surrounding environment to align with the visual aesthetic of the scene. Modifying the physical room’s appearance, such as walls or floors, to reflect the color palette and stylistic elements of the fictional world was believed to enhance users' sense of presence within that world. For example, adopting a saturated pink color scheme or a dilapidated, post-apocalyptic style could contribute to a stronger feeling of immersion.
Moreover, participants suggested adapting textures based on the architectural characteristics of the scene’s setting.  
For instance, we could incorporate elements resembling traditional Japanese or Gothic architecture.  
Such visual cues could help viewers more intuitively infer the geographical or cultural context of the scene.


\subsubsection*{\textbf{Character Presence (4/43)}}
Another key concept that emerged from participant feedback was the incorporation of film characters into the physical environment surrounding the viewer. For instance, in a circus-themed scene, populating the room with circus-related characters and performers was seen as a way to enhance immersion. By visually filling the space with such figures, users could more easily feel embedded within the narrative context. Participants also highlighted the importance of how characters enter the physical space. One example involved a fight scene in which the protagonist knocks down opponents, with the opponents being thrown out of the screen and into the room, effectively blurring the boundary between the screen and the physical space.

\subsubsection*{\textbf{Body Transformation (3/43)}}
In addition to populating the environment with characters, participants expressed interest in being able to embody the protagonist themselves. 
Specifically, altering parts of the user’s visible body, particularly the hands, to resemble those of the main character was considered a powerful technique for enhancing identification and immersion. For example, if the protagonist is a pirate, rendering the user’s hands with pirate-like features could strengthen the sense of embodiment within the story world.

\subsubsection*{\textbf{Window Augmentation (4/43)}}
Participants expressed interest in using windows in the physical environment. For example, several participants mentioned displaying fantastical worlds through the window, such as showing magical buildings in a fantasy film or a ruined town in a post-apocalyptic film. They see windows as a way to help viewers better understand and become immersed in the fictional world. Other participants proposed using windows to show recognizable real-world landmarks. For instance, they suggested displaying the Eiffel Tower, Big Ben, or the Statue of Liberty to signal that a scene is set in Paris, London, or New York, respectively. These augmentations were thought to aid viewers in understanding both the geographical setting and transitions between different locations in the film.

\subsubsection*{\textbf{Lighting Effects (5/43)}}
Participants also highlighted the use of lighting as a powerful means to reinforce the mood of a scene. For example, in horror films, participants suggested flickering lights or restricting illumination to a limited area of the room to amplify tension and fear. They also proposed incorporating context-specific light sources, such as moonlight streaming in or the headlights of cars, to enhance the atmosphere and narrative setting. Additionally, several participants emphasized the expressive use of shadows. One illustrative idea involved a dance floor surrounded by performers. Projecting the dancers' shadows around the viewer could evoke the feeling of being within that crowd.

\section{\system{}: SYSTEM DESIGN}
\subsection{Overview}
We present \system{}, a system that leverages generative AI to semantically extract key visual and contextual elements from film scenes, and automatically generates scene-related visual augmentations within the user's physical environment. By seamlessly integrating these elements into the real world, \system{} enhances viewer's immersion into the film.

\subsection{System Pipeline}
\begin{figure*}[t]
\centering
\includegraphics[width=\textwidth]{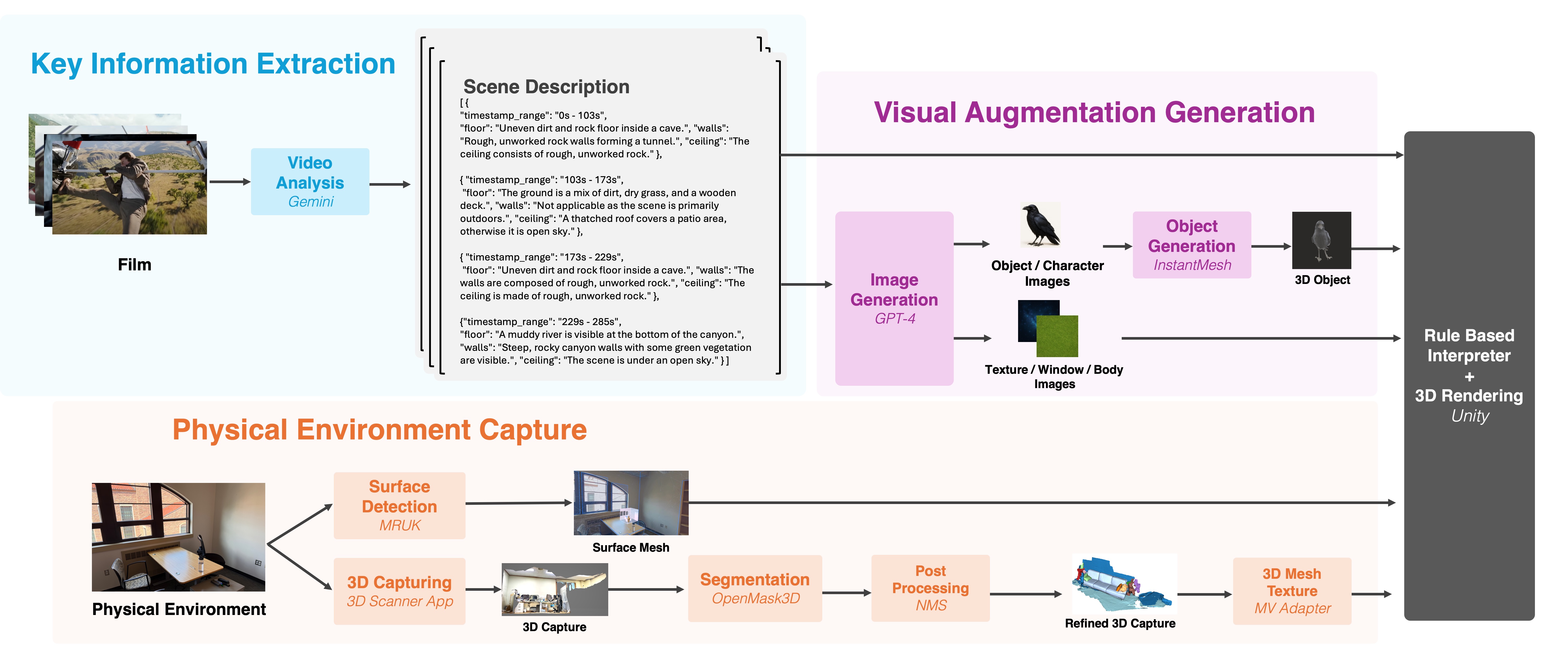}
\caption{
System pipeline of \system{}. Our pipeline consists of three key stages: 1) key information extraction, 2) visual augmentation generation, and 3) physical environment capturing.}
\label{fig:pipeline}
\end{figure*}

Our system pipeline is shown in Figure~\ref{fig:pipeline}. This pipeline consists of the following three stages: 1) extracting key semantic elements from film scenes, 2) capturing and analyzing the user’s 3D physical environment, and 3) generating and rendering scene-appropriate visual augmentations within the user’s surroundings.  

\subsubsection*{\textbf{Step1. Extracting Key Information from Films}}
The first step of our pipeline is to analyze the film. Our system processes the video file to extract timestamped semantic and visual information from each scene. To perform this analysis, we utilize the Gemini API (gemini-2.5-pro), which allows frame-level querying and contextual feature extraction based on large-scale visual-language models. Our system gives the Gemini API a movie file with MP4 that we want to augment, along with prompt text files, and receives JSON files as output. Our system uses Gemini to extract scene-level features across the following seven categories: 1) particle effects, 2) surrounding objects, 3) room textures, 4) character presence, 5) body transformation, 6) window augmentation, and 7) lighting effects. Our system uses Gemini independently for each feature, obtaining a separate JSON output for each. We prompted Gemini to divide the scene into segments where changes occur based on each feature. We asked Gemini to get the details of each feature for each segment, including the feature name, property (size, position), and context. We use a few-shot prompt by providing several example outputs, and the prompt text is customized based on each feature and is shown in the appendix.


\subsubsection*{\textbf{Step 2. Capturing and Analyzing Physical Environment}}
Our system supports two types of environment scanning: a lightweight "simple scanning" mode and a high-fidelity "detailed scanning" mode. This dual approach balances usability and precision, as detailed scanning requires additional hardware and longer processing time (about 30 minutes). For simple scanning, we utilize the Meta Quest 3’s built-in mixed reality capabilities, which can detect large planar surfaces such as walls, floors, ceilings, and windows using its surface detection toolkit (Mixed Reality Utility Kit). While fast and device-integrated, this mode provides only coarse spatial information and does not capture fine-grained object geometry or semantics.
The resulting data consists of a set of semantic labels and associated information, including position, size, and rotation. (e.g. name: 'floor', position: [x, y, z], size: [x, y, z], rotation: [pitch, yaw, roll]).

For detailed scanning, the environment is first pre-captured by users using the 3D Scanner App~\cite{3DScanner}, a mobile application that utilizes LiDAR sensors to generate high-resolution 3D meshes. The captured mesh is then segmented using OpenMask3D~\cite{takmaz2023openmask3d}, followed by refinement and outlier filtering via HDB-SCAN~\cite{Thehdbsc36}. Finally, textures are added using MVAdapter~\cite{huang2024mv}, which takes both the refined 3D mesh and a textual prompt extracted from the scene description JSON file.

\subsubsection*{\textbf{Step 3. Generating Visual Augmentations within the Physical Environment}}
Based on the extracted scene descriptors, which are stored as JSON files, our system generates corresponding textures and images to be placed within the user’s environment using image generation AI, ChatGPT 4o. For 3D objects, our system sends the generated images to the 3D object generation tool, InstantMesh~\cite{xu2024instantmesh}, with prompts from the scene descriptors and generates corresponding 3D objects.

These semantic analysis and visual augmentation generation happens offline for each film in advance, and then these film props and the data of the physical environment are transferred to Unity. In Unity, our system takes the following steps. First, it reads the scene descriptors and stores the information. Then, based on this information, the system displays the augmentations at the specified timing and with the specified spatial configuration (size and position). Each element of the physical environment (e.g., floors, windows) is saved as a GameObject in the Unity scene. Therefore, our system can place these GameObjects with the augmentations. For detailed 3D mesh, user place the 3D mesh to align the position to the space with their controllers. Details for each feature are shown in the next section.

Once the augmentations are generated, the system is ready for use. The user wears an MR headset (Quest 3), through which they can view the film on a virtual display in front of them, while the surrounding physical space is augmented with contextually relevant visual elements.

\subsection{Film Augmentation Features}
Based on the design space, we developed the following seven augmentation features: 1) particle effects, 2) surrounding objects, 3) room textures, 4) character presence, 5) body transformation, 6) window augmentation, 7) lighting effects.
We implemented this system with Unity (version 6000.1.5f1) and deployed it to Meta Quest 3. We used MRUK (version 76.0.1) for spatial understanding and the Depth API (version 72.0.0) for handling occlusion between real and virtual elements.

\subsubsection{\textbf{Particle Effects}}
\begin{figure}[h]
\centering
\includegraphics[width=\imgparam\linewidth]{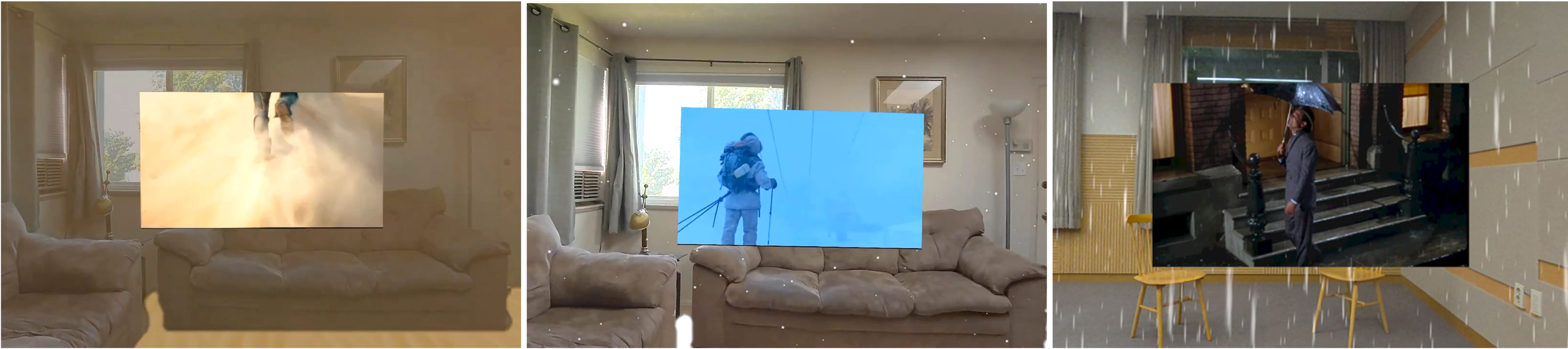}
\caption{Particle Effects Examples. 1) Dust, 2) Snow, 3) Rain.}
\label{fig:particles}
\end{figure}

Particle effects enhance immersion by introducing environmental effects such as rain and snow into the user's physical surroundings, as shown in Figure~\ref{fig:particles}. The particle type and behavior are selected by analyzing the scene descriptor in the JSON file. Our system interprets two attributes from the JSON: type and density. The system currently supports five types: rain, snow, dust, smoke, and fog. For density, there are four patterns: sparse, moderate, dense, and torrential. Each of these combinations has different custom particle animations.

For the particle effect, our system reads the entry `"type": \{particle name\}' and selects the corresponding particle effect. To create the five particle types, we used the Particle System module in Unity to visualize the particle effects. We manually set the position, texture image, and color for each particle type. All particle materials are assigned an unlit transparent shader so that they do not block other visual effects.

For density, our system reads the entry `"density": \{density\}' and adjusts the "Rate Over Time" parameter of the Particle System module accordingly. Different values are set for each particle type. The number of particles for \{sparse, moderate, dense, torrential\} is \{50, 100, 150, 250\} for rain, \{100, 200, 300, 600\} for snow, and \{4, 7, 10, 15\} for dust, smoke, and fog.

\subsubsection{\textbf{Surrounding Objects}}
\begin{figure}[h]
\centering
\includegraphics[width=\imgparam\linewidth]{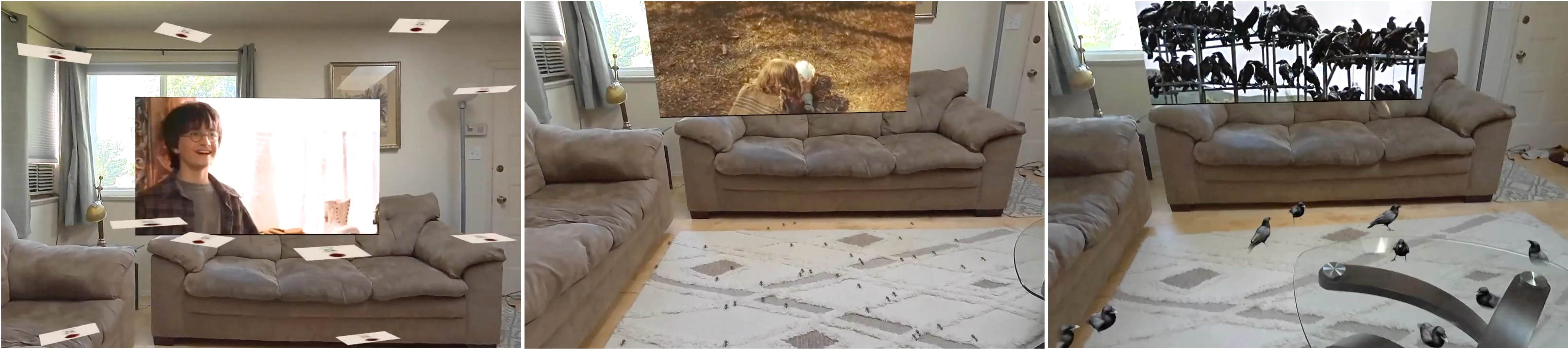}
\caption{Surrounding Objects Examples. 1) Letters in the air, 2) Bugs crawling on the floor, 3) Birds around.}
\label{fig:Objects}
\end{figure}

Surrounding objects bring objects from the film into the user’s environment, as shown in Figure~\ref{fig:Objects}. Based on object appearance descriptions, our system uses a text-to-image generation model (gpt-image-1) to produce reference visuals, which are then converted into 3D meshes using Instant Meshes. In addition, the scene descriptions specify attributes such as size, position, and number. Due to limitations in scene understanding, we approximately define the size of an ant as 0.02 and that of an average human as 1.0. To prevent objects from becoming overly large and obstructive within the user’s environment, we cap the maximum size ratio at 1.3. Position is categorized as either "on the floor" or "in the air," which corresponds to being placed directly on the floor or 50 cm above it, respectively. To prevent large objects from blocking the screen, if an object's size ratio exceeds 0.5, we place it either to the side or behind the user. For objects smaller than 0.5, we place them at random positions within a 2m × 2m space centered around the screen. The number of objects is determined by qualitative labels: "a few," "several," "many," and "countless." These labels are mapped to quantitative values as follows:  
"a few" = $2/\text{size\_ratio}$,  
"several" = $4/\text{size\_ratio}$,  
"many" = $10/\text{size\_ratio}$, and  
"countless" = $20/\text{size\_ratio}$.  
To accommodate the computational limitations of the Quest 3 headset, we cap the maximum number of rendered objects at 20.

To simulate movement, our system uses the Particle System module in Unity. Motion behaviors are defined by the "pattern" attribute within the "motion\_details" field of the scene description. We currently support five movement patterns: "away\_from\_static", "to\_user", "away\_from\_user", "chaotic", and "falling". For each pattern, we manually set the "Velocity over Lifetime" property in the Particle System across all axes. For example, in the "to\_user" pattern, the velocity is set to -0.5 along the z-axis (longitudinal axis toward the user), causing the particles to approach the user.

\subsubsection{\textbf{Room Textures}}
\begin{figure}[h]
\centering
\includegraphics[width=\imgparam\linewidth]{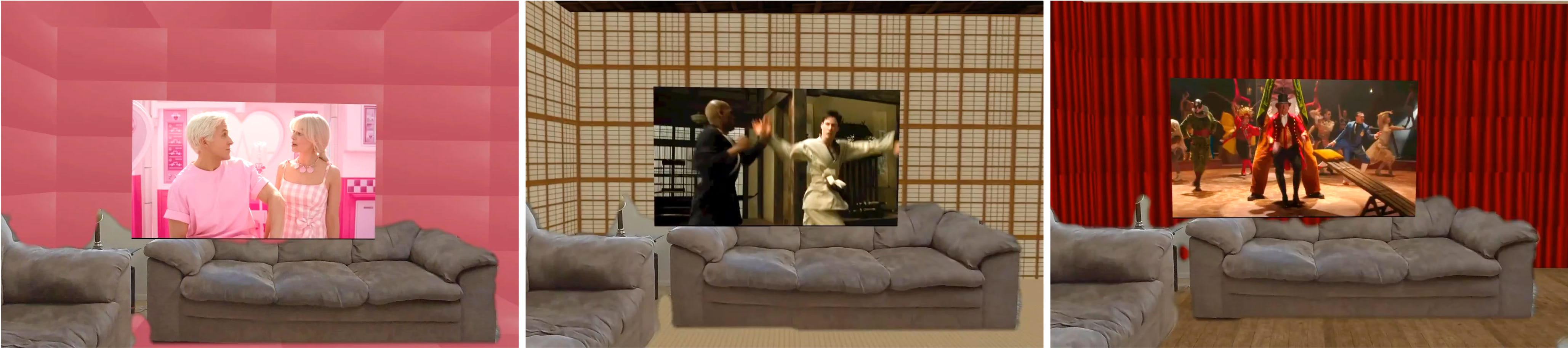}
\caption{Room Texture Examples. 1) Vivid-color environment, 2) Japanese-style environment, 3) Circus-like environment.}
\label{fig:Texture}
\end{figure}

Room textures adapt the visual appearance of the user’s environment to match the thematic setting of the film, as shown in Figure~\ref{fig:Texture}. Based on the scene descriptor in the JSON file, which includes a textual explanation of the room texture, the system generates texture images using a text-to-image model (gpt-image-1). To apply these textures to the 3D mesh of the environment, our system obtains the coordinates and dimensions of the real-world floor, ceiling, and wall surfaces. Our system then places pre-defined Quad objects at the corresponding locations. Each Quad is subdivided into a 5×5 grid, resulting in 25 tiles in total. The generated texture images are applied to each subdivided Quad based on its corresponding surface type (e.g., floor, wall, ceiling). To support lighting effects, which we describe later, we apply a "Lit" shader for all Quads. In addition, to avoid rendering conflicts such as z-fighting between the original mesh surfaces and the textured Quads, we use the Depth API’s MeshFilter component to occlude the original surfaces where new textures are overlaid.

\subsubsection{\textbf{Character Presence}}
\begin{figure}[h]
\centering
\includegraphics[width=\imgparam\linewidth]{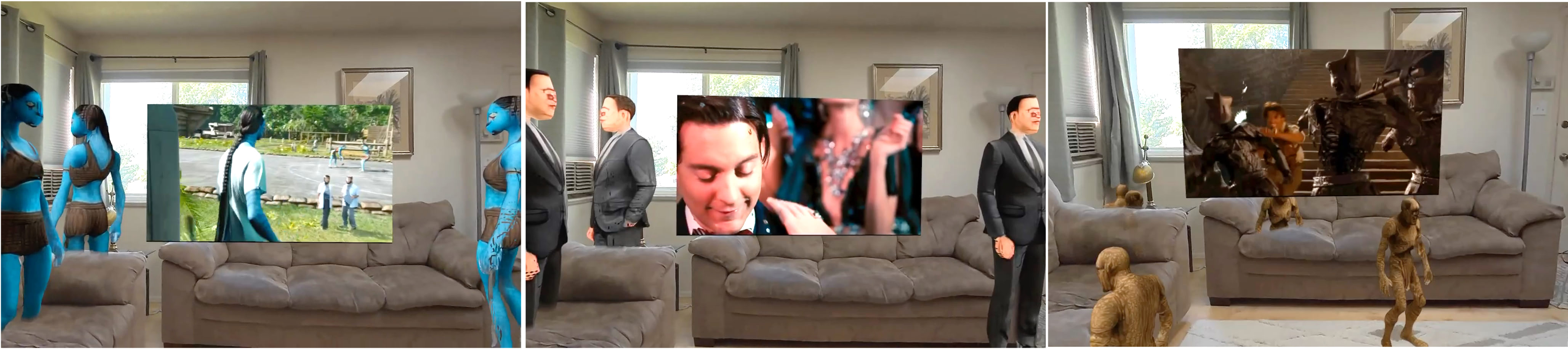}
\caption{Character Presence Examples. 1) Blue-creatures, 2) People in party, 3) Mummies.}
\label{fig:CharacterPresence}
\end{figure}

Character presence brings key characters from the film into the user’s physical environment, as shown in Figure~\ref{fig:CharacterPresence}.
To generate these characters, our system uses the scene’s JSON descriptor to extract relevant character attributes. A text-to-image model (gpt-image-1) generates visual references based on these descriptions. We found that square images (1024$\times$1024) often failed to capture the full body of a character. Therefore, we used a 1024$\times$1536 resolution to ensure that full-body depictions were included. The resulting images are then converted into 3D meshes using InstantMesh. The size ratio and position for character presence are handled in the same manner as for surrounding objects. Regarding motion, the scene descriptor classifies characters as either "static" or "moving". For characters labeled as "moving", the system programs them to move randomly along the x- and z-axes at speeds ranging from -0.5 to 0.5 m/s, while maintaining a fixed y-axis (vertical height).





\subsubsection{\textbf{Body Transformation}}
\begin{figure}[h]
\centering
\includegraphics[width=\imgparam\linewidth]{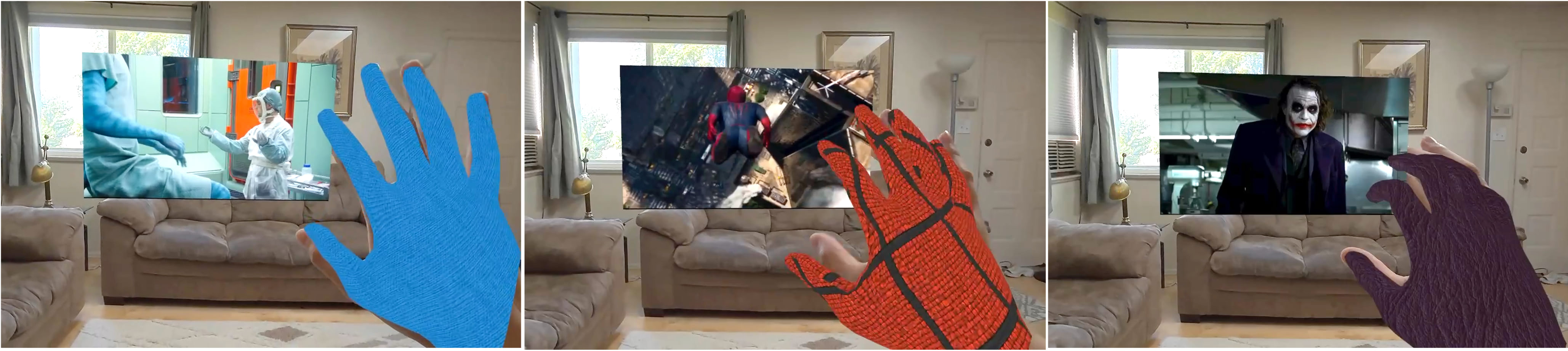}
\caption{Body Transformation Examples. 1) Fantasy Creature, 2) Heroic Costume, 3) Villain outfit .}
\label{fig:Body Transformation}
\end{figure}

Body transformation changes the user’s own virtual appearance to align with the film’s narrative, as shown in Figure ~\ref{fig:Body Transformation}. 
To generate these virtual appearance, our system gets the description of the appearence of the key character from the scene descriptor.
Our system then generates textures from the scene descriptors using a text-to-image model (gpt-image-1). Using the Quest 3’s Hand Tracking SDK, our system detects the user’s hands in real time, and the generated textures are dynamically overlaid to reflect the character’s appearance. 


\subsubsection{\textbf{Window Augmentation}}
\begin{figure}[h]
\centering
\includegraphics[width=\imgparam\linewidth]{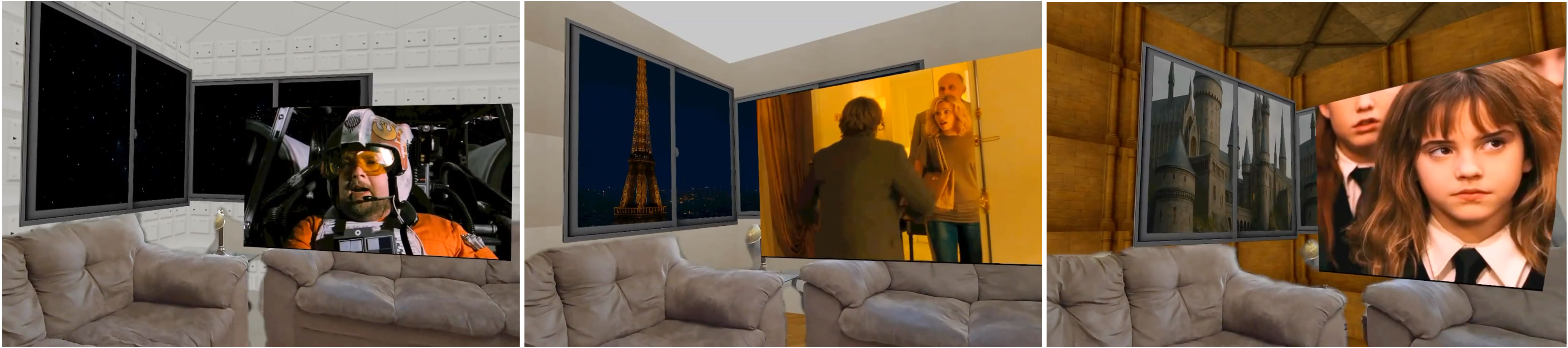}
\caption{Window Augmentation: 1) Surrounded space, 2) Landmark, 3) Key building. }
\label{fig:WindowAugmentation}
\end{figure}

Window augmentation displays an outdoor scene through the window to help users better understand the context of the film, as shown in Figure~\ref{fig:WindowAugmentation}.  
Based on the scene descriptor, our system first generates an image representing the view outside using a text-to-image model (gpt-image-1). Using MRUK, the system retrieves the 3D object prefab that defines the position and dimensions of the room’s window. The generated image is then rendered onto the window surface and overlaid with a custom-designed window frame image to simulate a realistic appearance. To ensure that the frame is rendered on top of the window image, it is placed 0.01 units forward along the window's forward direction. Additionally, to emphasize that it is an exterior view, we use an "Unlit" shader for the outdoor image, keeping it unaffected by scene lighting, in contrast to the "Lit" shaders used for room textures.

\subsubsection{\textbf{Lighting Effects}}
\begin{figure}[h]
\centering
\includegraphics[width=\imgparam\linewidth]{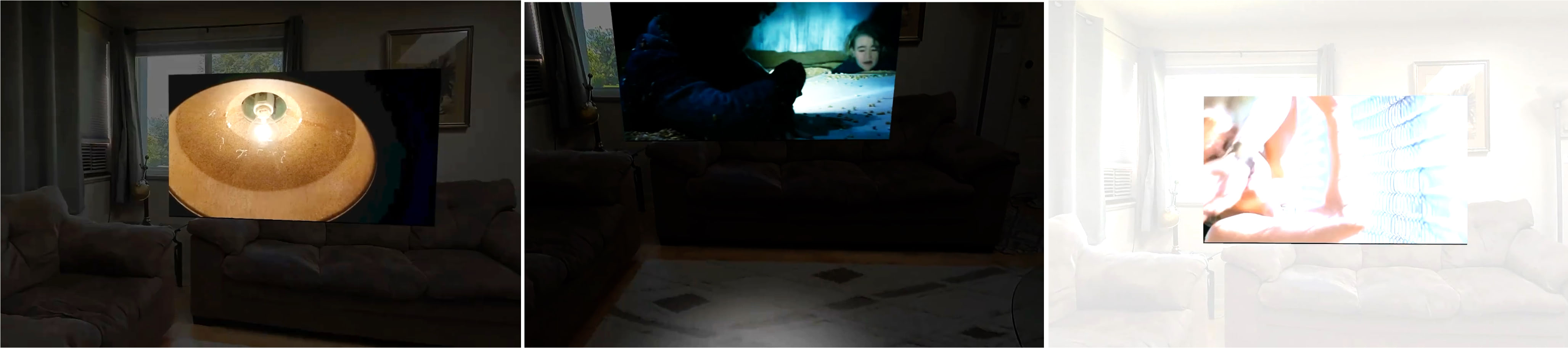}
\caption{Lighting Effects. 1) Low brightness / Flickering, 2) Spotlight, 3) Whiteout.}
\label{fig:LightingEffects}
\end{figure}

Lighting effects help to convey the mood of a scene, such as horror, suspense, or tension, as shown in Figure~\ref{fig:LightingEffects}.  
Based on the brightness level shown in the scene description, our system changes the brightness through MRUK's passthrough layer in Unity. To maintain visual consistency for the surfaces generated in the Room Textures module with the overall changes in scene illumination, the \textit{Ambient Light} value in Unity's lighting settings is also adjusted. There are five brightness levels: "very\_dark", "dark", "normal", "bright", and "overexposed". The brightness parameters of both MRUK's passthrough layer and the \textit{Ambient Light} are set to \{0.25, 0.5, 1, 2, 4\}, corresponding to the above levels.

Then, our system apply three types of lighting effects: flickering light, spotlight, and whiteout. Our system retrieves the type name from the "type" field in the JSON and enables the corresponding effect. Flickering light and whiteout effects modify the brightness of both the passthrough layer and the ambient light. The flickering light effect randomly multiplies the brightness of these sources by a factor between 0.4 and 2.0 over time, creating a fluctuating lighting pattern. In contrast, the whiteout effect multiplies the brightness value by 2 to simulate a strong, bright light across the entire scene. The spotlight effect is implemented using Unity's native Spot Light component, which projects a focused beam of light onto the floor beneath the 2D screen.

\subsection{Technical Evaluation}

\subsubsection{Method}
We conducted a technical evaluation to evaluate the accuracy and generalizability of our system. We randomly collected 100 movie clips from five genres (action, drama, musical, horror, and sci-fi) with 20 clips per genre, ensuring a diverse set of scenes for evaluation. Each clip had a duration of 3 to 5 minutes. Due to the lack of a standardized and automated evaluation method, two authors (A1 and A3) manually reviewed the system-generated results. 

We assessed the accuracy of the system by identifying errors based on all seven features of our system: 1) particle effects, 2) surrounding objects, 3) room textures, 4) character presence, 5) body transformation, 6) window augmentation, and 7) lighting effects. For analysis, we defined an error as any misalignment with the original content. Errors were classified into three categories: visual, semantic, and temporal. Visual misalignment refers to discrepancies in appearance, such as color, texture, brightness, position, and scale. Semantic misalignment refers to inconsistencies in meaning or context between the generated output and the original scene. Temporal misalignment refers to errors in timing, including effects, object or character appearance, and texture changes.

In addition, we measured the preprocessing time for each step of the pipeline, including video analysis, image generation, and object generation. We used a desktop PC (Intel i9 CPU, NVIDIA RTX 3090 GPU) for API communication, and a server (NVIDIA A100 GPU) for object generation.

\begin{figure}[h]
\centering
\includegraphics[width=\linewidth]{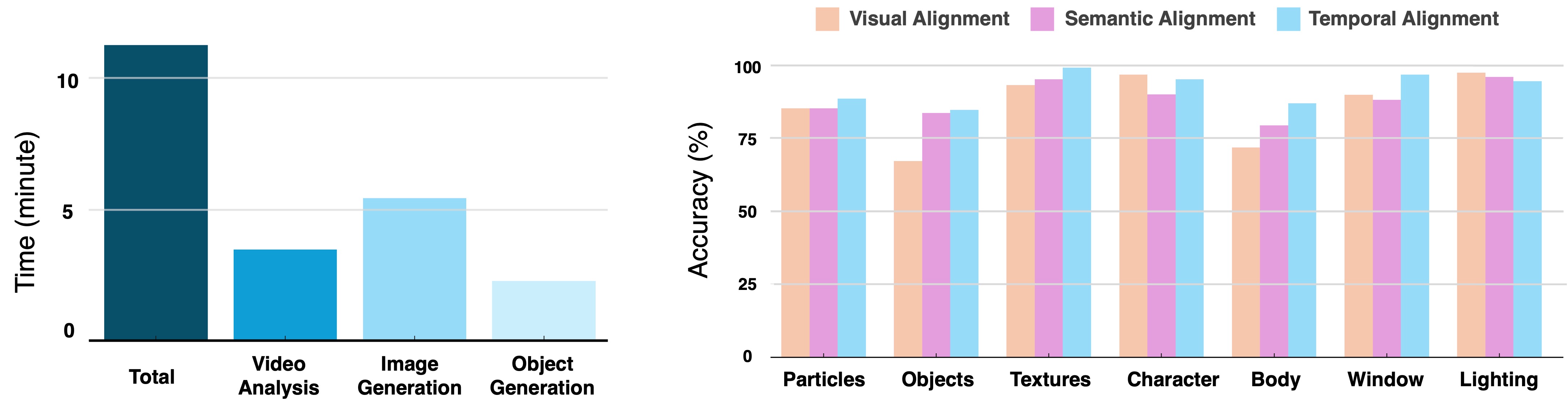}
\caption{Technical evaluation results. The left graph shows the results of the preprocessing time, including the total time and the time for each step (video analysis, image generation, and object generation) per video. The right graph shows the accuracy of visual, semantic, and temporal alignment for each augmentation type.}
\label{fig:techeval_results}
\end{figure}

\begin{figure}[h]
\centering
\includegraphics[width=\imgparam\linewidth]{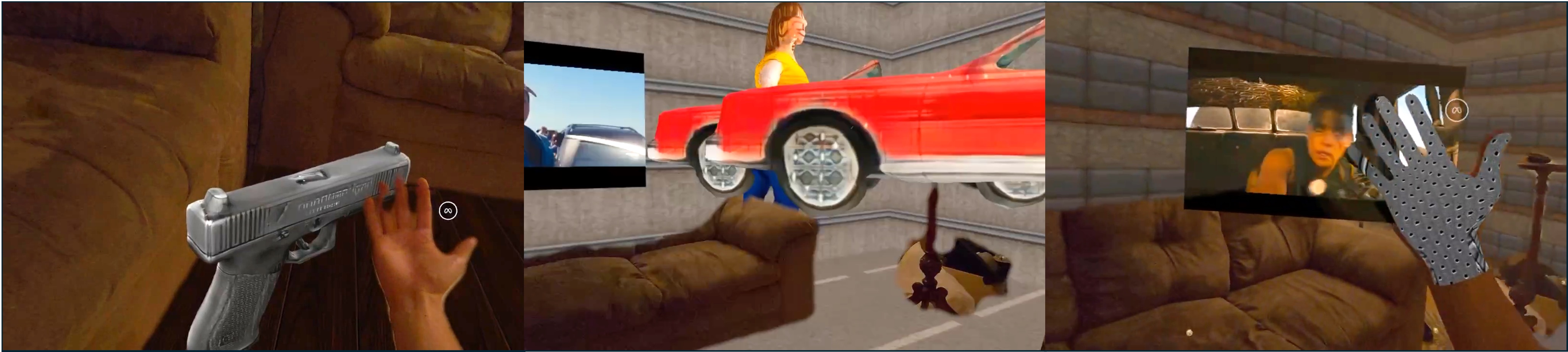}
\caption{Examples of typical errors identified in the technical evaluation. 1) A disproportionately large handgun, representing a visual alignment error in surrounding objects; 2) A floating car, also represents a visual alignment error in surrounding objects; 3) Inconsistent hand textures, representing a semantic alignment error in body transformation.}

\label{fig:techeval_examples}
\end{figure}
\subsubsection{Results}
Figure~\ref{fig:techeval_results} (left) shows the result of the preprocessing time. The mean time for each clip (3-5min) was 3.49 minute for video analysis, 5.44 minute for image generation, 2.30 minute for object generation. Although the total time was 11.24 minute per clip, all steps are processed in advance. Therefore, viewers do not experience any processing delays during playback once the preprocessing has been completed.

Figure~\ref{fig:techeval_results} (right) shows the results of the technical evaluation across 21 items (= 3 measurement item $\times$ 7 augmentation types). 18 of 21 items showed an accuracy above 80\%, indicating that our system performs reliably across most augmentation types. However, three items fell below this threshold: visual alignment for surrounding objects (67\%), visual alignment for body transformation (72\%), and semantic alignment for body transformation (79\%).

Figure~\ref{fig:techeval_examples} shows typical errors identified in the technical evaluation. For the visual alignment of surrounding objects, there were errors related to size, position, and brightness. Regarding size, there were cases of inappropriate object sizes, such as an excessively large handgun and an overly small sandbag. This issue may have been caused by errors in scene understanding. Regarding position, there were incorrect placements, such as floating cars and beer bottles stuck on the ground. This error may have been caused by both misinterpretation during video analysis and unexpected origin positions generated by InstantMesh. As for brightness, some objects were occasionally too bright, such as the sandbag and the bonsai tree, which appeared overly bright compared to their original appearance. This issue may be related to the current implementation, which uses a single material and shader for all objects.

For body transformation's semantic alignment, unrelated hand textures occasionally appeared in the scene, such as a white glove emerging from nowhere. This issue could have been caused by a misinterpretation during video analysis. In this case, the JSON output included the entry "user\_character": "white glove".
For body transformation's visual alignment, there were cases where the hand texture did not match well with the intended appearance. For example, while the expected texture was a realistic human hand partially covered with green and black smudges, the augmented output depicted a hand entirely green in color. This error could be attributed to a failure in the image generation process. In this case, the JSON clearly specified "user\_character": "hand covered in a green and black smudge", indicating that the image generation did not accurately reflect the input description.

\section{User Study}
We evaluated our system from two perspectives: one of the viewers and another of the creators.
First, we conducted a usability study with twelve participants to gather insights from the viewers’ perspective, comparing our system with no augmentation and with 2D augmentation approaches.
Second, we conducted expert interviews with eight film creators to explore how they perceive viewer use of our system and to collect suggestions for potential improvements.

\subsection{User Evaluation}
\subsubsection{Methods}
\begin{figure}[h]
\centering
\includegraphics[width=\imgparam\linewidth]{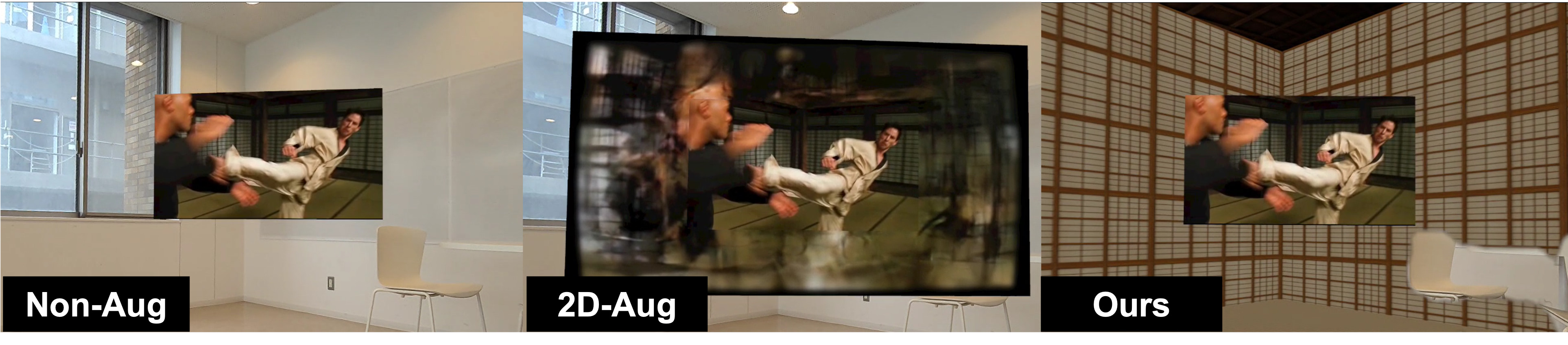}
\caption{Participant’s view in the three conditions.}
\label{fig:UserStudy_Conditions}
\end{figure}

We recruited 12 participants (7 male, 5 female; aged 21–25, M = 23.0, SD = 1.0) from our local university. 
Participants reported their interest in films (M = 5.5, SD = 1.2) on a 7-point Likert scale (1 = Not interested at all, 7 = Very interested), and their self-reported familiarity with MR devices (M = 2.8, SD = 1.6; 1 = Not familiar at all, 7 = Very familiar).
After obtaining informed consent and explaining the research objectives, participants wore the MR headset (Meta Quest 3) and were allowed to freely look around to familiarize themselves with the environment.
Participants then experienced the following three conditions: 
\begin{itemize}
  \item \textbf{[\textit{Non-Aug}] Non-augmentation:} A 2D screen was displayed in front of the user without any additional augmentation.
  \item \textbf{[\textit{2D-Aug}] 2D-augmentation:} A 2D screen was displayed in front of the user, with its periphery augmented using a pix2pix-based method following \textit{ExtVision}~\cite{kimura2018extvision}. Specifically, we adopted their \textit{Recursive Method}, in which the same video is used to create the training dataset by pairing cropped and original frames. We trained for 200 epochs with a patch size of 256×256. In addition, we applied deflickering, lens distortion, and Gaussian blur on edges, based on the paper.
  \item \textbf{[\textit{Ours}] \system{}:} A 2D screen was displayed in front of the user, with our augmentation rendered in the surrounding environment.
\end{itemize}
The participant’s view for each condition is shown in Figure~\ref{fig:UserStudy_Conditions}.

We used 21 clips from films covering a broad range of genres, including adventure, horror, science fiction, suspense, comedy, musical, and action, with three clips highlighting each of the seven augmentation types. The total duration is approximately 10 minutes, with each clip lasting approximately 30 seconds, and the list of films is shown in the appendix.
In all conditions, participants viewed the same set of 21 clips. We used a within-subjects design in which each participant experienced all three conditions. To counterbalance order effects, we employed all six possible permutations of the three conditions, assigning two participants to each order. 

\begin{figure}[h]
\centering
\includegraphics[width=\imgparam\linewidth]{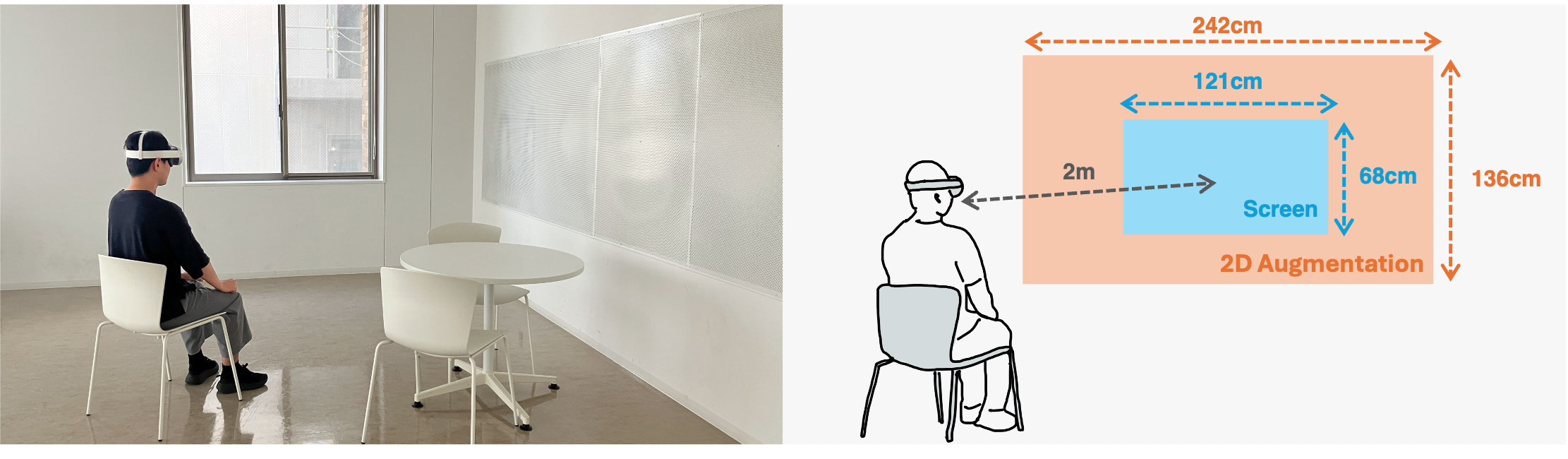}
\caption{User Study Settings. Environment (left) and window position/size (right).}
\label{fig:UserStudy_Settings}
\end{figure}

The environment of this study is shown in Figure~\ref{fig:UserStudy_Settings}. 
The experiment was conducted in an area approximately 4.1\,m (width) $\times$ 4.1\,m (depth) $\times$ 2.9\,m (height). The room was furnished with two chairs, one desk, and one window. We used a 55-inch virtual screen (121 cm wide × 68 cm high) positioned 2 m from the viewing position, based on previous research~\cite{jones2013illumiroom,kimura2018extvision} and commercial product guidelines~\cite{ComfortM63:online}. 

After each condition, participants filled out a 7-point Likert scale questionnaire (1 = strongly disagree, 7 = strongly agree) assessing the following aspects, based on previous HCI studies~\cite{ryu2021gamesbond,je2019aero,kimura2018extvision} and additional custom questions:
\begin{itemize}
  \item \textbf{Immersiveness:} ``I felt as if I was inside the world of the film.''
  \item \textbf{Enjoyment:} ``I enjoyed the experience.''
  \item \textbf{Consistency:} ``I felt that the film and the surrounding environment were connected.''
  \item \textbf{Fatigue:} ``I felt tired during the experience.''
  \item \textbf{Distraction:} ``The augmentation distracted me from the film.''
  \item \textbf{Regular use:} ``I would like to use this regularly when watching films.''
  \item \textbf{Film–augmentation match:} ``The augmentation felt well-matched to the film content.''
  \item \textbf{Focus:} ``I was able to stay focused on the film during the experience.''
\end{itemize}
After all conditions, participants also rated their satisfaction with each feature of \system{} individually using a 7-point Likert scale questionnaire (1 = strongly disagree, 7 = strongly agree). After completing all conditions, we conducted semi-structured interviews to gather in-depth qualitative feedback.  All sessions were conducted in person, lasted approximately 60 minutes, and participants received \$15 as compensation.

\begin{figure*}[t]
\centering
\includegraphics[width=\textwidth]{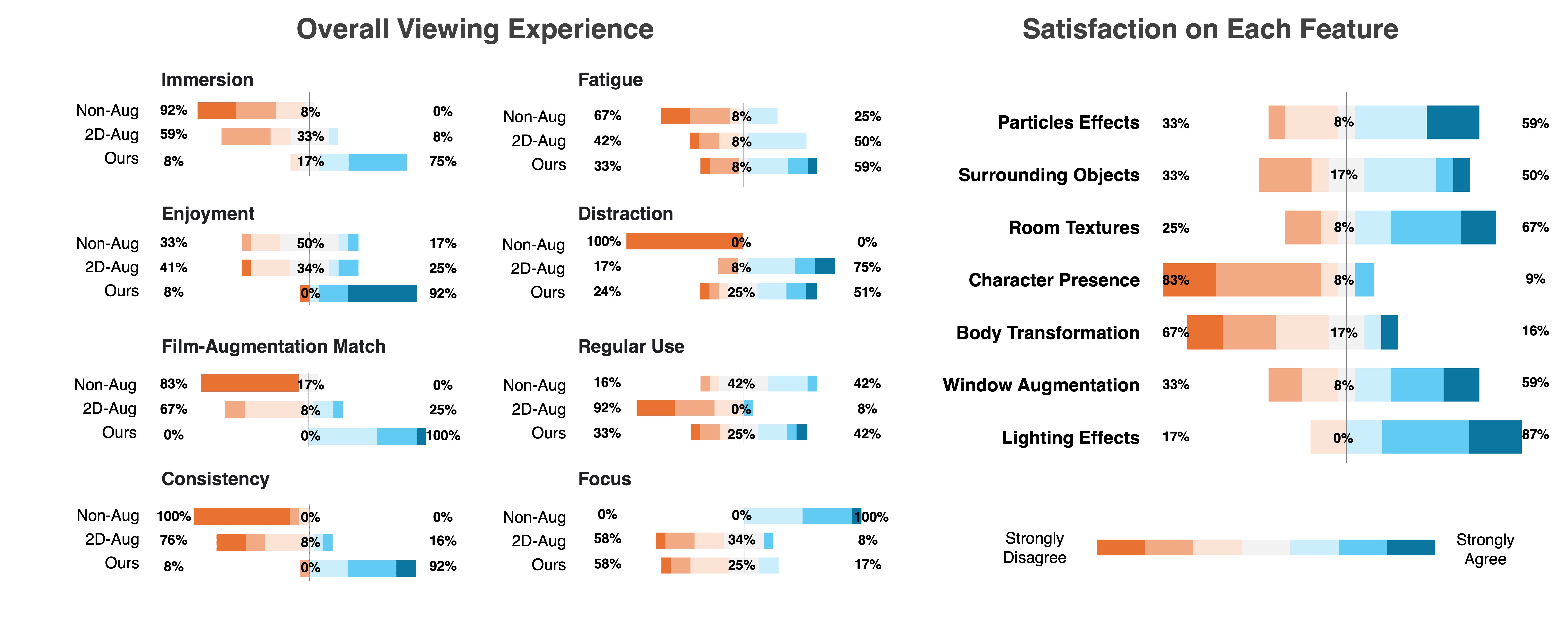}
\caption{
Results of the User Study. The left graph is the rating of the overall viewing experience, and the right graph is the rating of the satisfaction on each feature.}
\label{fig:userstudy_results}
\end{figure*}
\subsubsection{Results}
Figure~\ref{fig:userstudy_results} shows the comparison of the three conditions (\textit{Non-Aug}, \textit{2D-Aug}, \textit{Ours}) and the result of the satisfaction on each feature.

\subsubsection*{\textbf{1) Overall \system{} Viewing Experience}}

\textit{Ours} was rated as the most immersive and enjoyable among the three conditions. In terms of \textbf{Immersion}, 75\% of participants agreed or strongly agreed that \textit{Ours} provided an immersive experience, compared to 8\% for \textit{2D-Aug} and 0\% for \textit{Non-Aug}. Participants described feeling deeply absorbed, with one stating, \textit{“I felt as if I were in the same environment as the protagonist”} (P5), and another commenting, \textit{“I liked how this system removed the sense of the real world around me”} (P3). Regarding \textbf{Enjoyment}, 92\% of participants reported enjoyment with \textit{Ours}, compared to 25\% for \textit{2D-Aug} and 17\% for \textit{Non-Aug}. Participants emphasized how the system enhanced their emotional engagement. For instance, P10 noted, \textit{“It was fun to feel like I was actually inside the scene,”} and P2 mentioned, \textit{“It felt like what was happening in the movie was really happening, and I was experiencing it myself.”} Based on functional aspects, participants also rated \textit{Ours} positively. Regarding \textbf{Film–Augmentation Match}, all participants (100\%) agreed or strongly agreed that the augmentation felt well-matched to the film content, compared to only 25\% for \textit{2D-Aug}. In terms of \textbf{Consistency}, 92\% of participants felt that the film and the surrounding environment were connected in \textit{Ours}, whereas only 16\% reported the same for \textit{2D-Aug}, suggesting that our system provided a more coherent and integrated viewing experience.

However, both \textit{2D-Aug} and \textit{Ours} received high scores for \textbf{Distraction} (\textit{2D-Aug}: 75\%, \textit{Ours}: 51\%),  and correspondingly low scores for \textbf{Focus} (\textit{2D-Aug}: 7\%, \textit{Ours}: 17\%). Some participants noted that augmentations appearing in front of the screen were distracting. For example, P1 stated, \textit{“Rain and sand weren’t too distracting, but objects that covered the screen were.”} Others mentioned being overly drawn to the augmented elements near the screen. P3 commented, \textit{“Because the content appeared close to the screen, I tried to look at the details and ended up getting distracted.”} This increased distraction may have contributed to the relatively high \textbf{Fatigue} score for \textit{Ours} (59\%).

Interestingly, participants were divided on the question of \textbf{Regular Use}. While 42\% expressed willingness to use the system regularly, 33\% disagreed. Several participants noted that our system may not be suitable for all film genres or viewing contexts. For instance, P8 remarked, \textit{“For serious or slow-paced films, where I want to focus on the content itself, strong visual augmentations don’t feel appropriate.”} Similarly, P5 stated, \textit{“I’d want to use it when I want to have a fun experience, but not when I want to concentrate on the film content.”}

\subsubsection*{\textbf{2) Each Feature Satisfaction}}

Overall, five features (particles, objects, textures, windows, and lighting) were generally positively received, with satisfaction rates of 59\%, 50\%, 67\%, 59\%, and 87\% respectively. In contrast, character presence (9\%) and body transformation (18\%) were rated less positively.

For \textbf{character presence}, participants expressed concerns about the unnaturalness and mismatch of the character’s movements relative to the film content. For example, P1 noted, \textit{“The character’s movements felt unnatural.”} Similarly, P8 commented, \textit{“I wanted the characters to move in a way that matched the video.”} P9 added a more specific expectation: \textit{“Since the protagonist was being attacked in the video, I thought the character should also attack me.”} Interestingly, while participants were critical of the character animation, they were more forgiving of other elements. As P1 stated, \textit{“The unnatural movement bothered me more than the unnatural textures.”} This suggests that expectations for animated characters may be higher than for other augmented elements like particles or textures.

For the \textbf{body transformation} feature, participants had mixed responses. Some found it difficult to engage with both their virtual hands and the movie simultaneously. P9 noted, \textit{“Because I was looking at my hands, I couldn’t really focus on the video, so it felt off.”} Similarly, P12 stated, \textit{“When I’m watching a movie, I don’t look at my hands.”}  However, others appreciated the feature and felt that it enhanced their sense of embodiment. P2 remarked, \textit{“I felt as if I had become the protagonist. I even wanted to make the same hand movements as Spider-Man, and actually did so.”} P7 added, \textit{“It didn’t feel like I was just watching. It felt like I was one of the characters.”}

Regarding \textbf{lighting effects}, this feature was one of the most positively received. Several participants mentioned that the dynamic lighting enhanced immersion and emotional engagement. For example, P12 shared, \textit{“With the added lighting, it really felt like I was in the scene.”} P4 described, \textit{“In the horror movie, the flickering lights created a sense of tension and made my heart race.”}

For the \textbf{room texture} feature, participants commented on how it influenced their perception of space, despite no change in the actual room layout. P11 explained, \textit{“When the texture changed to something like a spaceship interior, I felt a sense of confinement.”}

\subsection{Expert Interviews}
We conducted semi-structured interviews with eight filmmakers to collect expert feedback on our system. All participants had at least one year of experience in filmmaking, with an average of  8.25 years (ranging from 1 to 35 years). The objective of these interviews was to gain in-depth insights into filmmakers’ perspectives on our system.

\begin{table}[h]
\centering
\begin{tabular}{l l r}
\toprule
\textbf{Expert} & \textbf{Film Genre} & \textbf{Experience} \\
\midrule
E1 & Drama              & 3 \\
E2 & Drama              & 1 \\
E3 & Experimental Film  & 3 \\
E4 & Experimental Film  & 2 \\
E5 & Fantasy  / Crime    & 5 \\
E6 & Drama / Horror    & 7 \\
E7 & Drama / Horror    & 10 \\
E8 & Documentary   & 35 \\
\bottomrule
\end{tabular}
\end{table}

We asked the following questions:
\begin{itemize}
  \item \textbf{Q1:} What are the positive and negative aspects of using \system{}?
  \item \textbf{Q2:} What kind of films are suitable for using \system{}?
  \item \textbf{Q3:} What are filmmakers’ attitudes toward \system{} being used to their films?
  \item \textbf{Q4:} How do filmmakers envision their future creative workflows where audiences may view films through \system{}?
\end{itemize}
Each interview lasted approximately 60 minutes. Participants received \$20 USD as compensation.

\subsubsection{Method}
The study was conducted on-site. First, we asked participants about their filmmaking backgrounds and previous works, which took approximately 10 minutes. Next, we introduced the concept of our system and asked them to try the system implemented on a Meta Quest 3 headset. This usage phase lasted about 10 minutes. We presented several short film clips to demonstrate each key feature of the system. Finally, following the usage session, we conducted a semi-structured interview lasting approximately 40 minutes.

\subsubsection{Results}
\subsubsection*{\textbf{Overall Responses}}
All filmmakers (E1–E8) appreciated the \system{} experience, describing it as enjoyable, immersive, and dynamic. Four participants (E1, E6, E7, E8) explicitly stated they would be happy to apply it to their own work. They noted that the system was particularly well-suited for genres such as action, horror, and musical. While participants highlighted clear benefits, they also raised concerns and questions, pointing to future opportunities for refinement.  

\subsubsection*{\textbf{Immersion: Bringing Viewers Into the Film World}}
Most experts emphasized that \system{} significantly enhanced immersion, allowing viewers to feel as though they were inside the film—an effect previously unattainable. For horror and suspense films, augmentation was seen as heightening the sense of being in the same space as characters, intensifying emotional impact. E8 noted musicals as a strong fit, as augmentation made them feel as though they were on stage. E6 and E7 highlighted lighting effects as particularly powerful, stressing that lighting is one of the most technically challenging aspects of filmmaking. They felt that being able to manipulate environmental lighting could greatly aid in establishing mood.

\subsubsection*{\textbf{Atmosphere: Enhancing Cultural and Historical Understanding}}
Beyond immersion, participants suggested that our system could help audiences better grasp cultural or historical contexts. E1, E3, and E8 viewed documentaries as a promising application. For example, E3 noted that older documentaries often depict eras whose atmosphere is difficult for modern audiences to imagine. Augmentation could recreate that atmosphere, aiding comprehension. Similarly, E6 reflected on their own film set in the 1960s and remarked that adapting the environment to reflect the period’s mood would have been highly valuable.  

\subsubsection*{\textbf{Context: Conveying Background Information}}
Participants noted that \system{} could serve as a medium for conveying background information that may not be fully presented through the film alone. For instance, E8, who had produced a documentary on poverty in South Asia, described the difficulty of conveying not only people’s lives but also the environments shaping them. They felt that augmenting windows or room textures could supply valuable contextual cues about cities and interiors, providing audiences with a deeper understanding of the story.  

\subsubsection*{\textbf{Distraction: Risks of Over-Attention}}
Participants expressed mixed views on the universal benefits of augmentation. While E2 and E5 praised the particle effects as beautiful and engaging, E2 pointed out that they sometimes obstructed the screen, especially when appearing in the foreground. They suggested giving users the option to toggle whether augmentation occupied the foreground.  

\subsubsection*{\textbf{Fidelity: Risks of Altering Original Intent}}
Several experts warned that automated generation could unintentionally distort the filmmaker’s intent. E2, E6, and E7 noted that in documentaries, augmentation might skew facts or reduce objectivity. E2 compared documentary filmmaking to museum exhibitions, where maintaining distance and neutrality is crucial. E3 also worried about augmentations conveying unintended meanings. E2 was particularly resistant, stating they would only allow augmentation if they could design it manually.  

\subsubsection*{\textbf{Suitability by Genre and Visual Complexity}}
Participants emphasized that suitability depends on both genre and visual style. E2 and E3 suggested action films, particularly those featuring heroic protagonists, would be a strong match, while E8 argued that suspense films might not be appropriate, since their tension often relies on a restricted field of view. E1 noted that augmentation works best with visually complex, big-budget productions that use varied cuts, effects, and camera angles. In contrast, minimalist films with simple camera work or compositions could clash with augmentation, creating an unintended mismatch.  

\subsubsection*{\textbf{Possible Improvements}}
Three experts (E1, E2, E3) valued the character-focused augmentations, especially those applied to viewers’ bodies. They expressed interest in extending these effects, for instance by augmenting the legs or adding fantastical elements such as Spider-Man’s web-shooting. E4 suggested randomizing augmentation by character, allowing users to rewatch films from multiple perspectives. Participants also discussed ways to reduce cognitive load. E4 recommended lowering brightness or reducing the number of effects, while E2 and E3 suggested showing augmentation only at key points rather than continuously. E2 proposed a two-step experience: first watching the film without augmentation, then rewatching with augmentation to gain a fresh perspective.  

\subsubsection*{\textbf{Creator Control and Opportunities for Co-Creation}}
Experts emphasized the importance of giving creators control over augmentation. E5 preferred to design all augmentation themselves, while E2 wanted to control placement. E1 suggested limiting the number of shooting locations to keep the augmented world consistent and immersive. In contrast, E4 proposed opening augmentation to user co-creation, likening it to “mod” culture in gaming where fans expand storylines and features. This points toward a continuum—from fully automated augmentation, to user-driven co-creation, to creator-led control—highlighting diverse possibilities for integrating \system{} into filmmaking.

\section{Limitations and Future Work}
\subsubsection*{\textbf{Expanding to Other Modalities}}
Although our system presents various visualizations to augment films, it does not currently incorporate other modalities such as sound or physical feedback. Future implementations could explore additional modalities, including haptic feedback and audio-based feature extraction, to further enrich the viewing experience. For example, in an action scene where the ground shakes, physical movement could be conveyed to the viewer’s entire body through motion platforms. Similarly, in scenes set in foreign countries, adjusting the voices of surrounding characters to match the language spoken in the film could enhance both immersion and contextual consistency.

\subsubsection*{\textbf{Bringing the Real World into the Film}}
In this work, we augmented the film-viewing experience by extracting key information and visualizing the surrounding environment based on that information. However, this process could also be reversed so that information from the real world is extracted and integrated directly into the film content. For example, incorporating real-time weather data from the viewer’s environment into the film could enhance immersion, making the viewer feel as though they were part of the same world depicted in the film.

\subsubsection*{\textbf{Expanding to Different Domains}}
Although our focus was on enhancing the film-viewing experience, the proposed approach could be extended to other domains. One such example is Zoom-like 2D remote communication. Similar to \textit{BlendScape}~\cite{rajaram2024blendscape}, the system could extract contextual information from another participant’s environment and display key elements around the screen, integrating them into the user’s space. This approach could foster a stronger sense of co-presence as though they were sharing the same physical room. Another potential application is in sports viewing or instruction, where relevant information from a live game or training environment could be extracted and presented in the viewer’s surroundings to enhance both presence and immersion.

\section{Conclusion}
We introduced \system{}, a generative augmented reality system that augments viewers’ physical surroundings with automatically generated 3D mixed reality content extracted from 2D movie scenes. Through a formative elicitation study, we explored how physical environments can be augmented based on a film’s visual elements and identified seven key augmentation methods: particle effects, surrounding objects, room textures, character presence, body transformation, window augmentation, and lighting effects. Building on this design space, we proposed a novel film-to-MR augmentation pipeline. In a user study, we compared our approach against both a non-augmentation and a 2D-augmentation method, showing that our system can enhance immersion and enjoyment, while also introducing potential distraction and fatigue. Finally, expert interviews highlighted the importance of providing creators with greater control and revealed opportunities for collaborative co-creation among users.

\ifdouble
  \balance
\fi
\bibliographystyle{ACM-Reference-Format}
\bibliography{references}

\section*{Film Materials and Appendix}

We used images and video excerpts from the following films under the fair use policy for academic and research purposes:
\subsection{Films Used in the Figures}
\begin{itemize}
  \item Singin' in the Rain
  \item The Birds
  \item Star Wars: Episode IV -- A New Hope
  \item Indiana Jones and the Raiders of the Lost Ark
  \item The Terminator 2: Judgment Day
  \item The Matrix
  \item Harry Potter and the Sorcerer's Stone
  \item Harry Potter and the Chamber of Secrets
  \item The Mummy
  \item Spider-Man
  \item Avatar
  \item The Great Gatsby
  \item Midnight in Paris
  \item The Greatest Showman
  \item Annabelle: Creation
  \item A Quiet Place
  \item The Dark Knight
  \item Dune
  \item Barbie
\end{itemize}

\subsection{Films Used in the Formative Study}
\begin{itemize}
  \item \textit{Mission: Impossible – Dead Reckoning}
  \item \textit{Indiana Jones and the Temple of Doom}
  \item \textit{Mission: Impossible – Ghost Protocol}
  \item \textit{Harry Potter and the Prisoner of Azkaban}
  \item \textit{The Wild Bunch}
  \item \textit{Godzilla: King of the Monsters}
  \item \textit{John Wick}
  \item \textit{Gold Diggers of 1935}
  \item \textit{Damnation}
  \item \textit{Sherlock Holmes} (2009)
  \item \textit{Memoirs of a Geisha}
  \item \textit{Interstellar}
  \item \textit{Blade Runner 2049}
  \item \textit{Blade Runner}
  \item \textit{The Greatest Showman}
  \item \textit{The Nun II}
  \item \textit{Indiana Jones and the Kingdom of the Crystal Skull}
  \item \textit{The Devil Wears Prada}
  \item \textit{Star Wars: Episode I – The Phantom Menace}
  \item \textit{The Great Gatsby}
  \item \textit{Pirates of the Caribbean: The Curse of the Black Pearl}
  \item \textit{Barbie}
  \item \textit{Fear and Loathing in Las Vegas}
  \item \textit{Kingsman: The Secret Service}
  \item \textit{Kick-Ass}
  \item \textit{X-Men: Days of Future Past}
  \item \textit{Backdraft}
  \item \textit{The Ring}
  \item \textit{Inception}
  \item \textit{The Birds}
\end{itemize}

\subsection{Films Used in the User Study}
\begin{itemize}
  \item Singin' in the Rain
  \item The Birds
  \item Star Wars: Episode IV -- A New Hope
  \item Indiana Jones and the Raiders of the Lost Ark
  \item The Terminator 2: Judgment Day
  \item The Matrix
  \item Harry Potter and the Sorcerer's Stone
  \item The Mummy
  \item Spider-Man
  \item Avatar
  \item The Great Gatsby
  \item Midnight in Paris
  \item The Greatest Showman
  \item Annabelle: Creation
  \item A Quiet Place
  \item The Dark Knight
  \item Dune
  \item Barbie
\end{itemize}

\subsection{Prompts}
\subsubsection{Video Analysis Prompts} \textit{Common Rules.}
\begin{lstlisting}[label={lst:common_json}]
Common Rules for All Analyses:
1.  Segment Definition (Dynamic Segmentation): Analyze the entire video to identify visually distinct sections. A new segment is defined by a significant change in lighting, color, atmospheric effects, surrounding objects, surrounding characters, environment, or textures.
2.  Segment Filtering Rules:
    - Effects Rule: Prioritize segments with distinct visual effects. If none, analyze based on texture and context.
3.  Global Rule (Singular Form): All descriptive string values in the JSON output must be in the singular form (e.g., 'leaf', not 'leaves').
4. Global Rule (Timestamp Format): The timestamp format is mandatory. All timestamps must be represented as a total number of seconds from the start of the video.
Calculation: The total seconds are calculated as (minutes * 60) + seconds.
Data Type: Use integers or floating-point numbers (e.g., 72 or 108.2).
Example 1: A segment from 1 minute 12 seconds to 1 minute 30 seconds MUST be written as:
"timestamp_start": 72,
"timestamp_end": 90
5. Global Rule (Non-Overlapping and Prioritization): Each time segment (timestamp_start to timestamp_end) must contain only one visual effect for each feature type. If a feature analysis identifies multiple distinct visual effects for the same time segment (e.g., two different types of characters), you must follow these prioritization steps:
Identify and describe only the most prominent or visually dominant visual effect.
If prominence is equal, describe the visual effect that is most numerous.
Example: If a scene contains "small rock" and "large boulder" as surrounding objects, you must choose only one to describe. Based on the rules, the "large boulder" should be chosen as they are more visually prominent, even if less numerous. Do NOT create two separate entries for the same feature type within the same timestamp.
\end{lstlisting}

\subsubsection{Video Analysis Prompts} \textit{Particle Effects.}
\begin{lstlisting}[label={lst:particles_json}]
JSON Output Structure:
[
  {
    "timestamp_start": <number>,
    "timestamp_end": <number>,
    "atmospheric_particle": {
      "type": "[rain|snow|dust|smoke|fog]",
      "density": "[sparse|moderate|dense|torrential]"
    }
  }
]

Module-Specific Rules:
- Absence of Features: If no distinct atmospheric particles are present in the segment, the value for the `type` field should be `null`, and other fields within `atmospheric_particle` should be omitted or set to `null`.
\end{lstlisting}

\subsubsection{Video Analysis Prompts} \textit{Surrounding Objects.}
\begin{lstlisting}[label={lst:objects_json}]
JSON Output Structure:
[
  {
    "timestamp_start": <number>,
    "timestamp_end": <number>,
    "surrounding_object": {
      "description": "[A description of a single instance of a key, recurring object. e.g., 'torn paper', 'black feather', 'gold coin']",
      "position": "[on_the_floor|in_the_air]",
      "quantity": "[a_few|several|many|countless]",
      "size": "Given that an ant's size is 0.02 and a normal human's size is 1, return the expected size.",
      "motion_details": {
        "pattern": "[static|to_user|away_from_user|chaotic|falling]"
      }
    }
  }
]

Module-Specific Rules:
- Absence of Features: If there is no recurring object that fits the criteria, the value for `description` must be `null`, and other fields should be omitted.
- Recurring Elements: You must select an object that is recurring in multiple instances and defines the scene's environment.
- AI Policy Compliance: In `surrounding_object.description`, avoid depicting graphic or sensitive content. Replace it with a neutral alternative (e.g., "mutilated body part" becomes "twisted dark root").
- Completeness Rule: If `surrounding_object.description` is filled with a valid value, then all other fields within `surrounding_object` (position, quantity, size, motion_details) must also be filled with valid values.
\end{lstlisting}

\subsubsection{Video Analysis Prompts}\textit{Room Textures \& Window.}
\begin{lstlisting}[label={lst:context_json}]
JSON Output Structure:
[
  {
    "timestamp_start": <number>,
    "timestamp_end": <number>,
    "Contexts": {
      "overall_scene_context": "[brief description of overall scene atmosphere and vibe. e.g. rainy late night outdoor in paris]",
      "outside_view_context": "[if the scene is indoor, describe the representative outside view scene. e.g. futuristic city skyline through a window]",
      "floor_texture_keywords": ["keyword1", "keyword2", ...],
      "wall_texture_keywords": ["keyword1", "keyword2", ...],
      "ceiling_texture_keywords": ["keyword1", "keyword2", ...]
    }
  }
]

Module-Specific Rules:
- Absence of Features: If an element for a given field is absent, represent string values as `null` and lists as an empty list `[]`.
- Image Generation-Friendly Keywords: All keywords and description in `Contexts` must be visually concrete and specific (e.g., 'rotting plank,' 'brushed steel'). Do not use subjective or abstract terms like 'beautiful' or 'scary'.
- Content Policy Compliance: All keywords and descriptions must adhere to standard AI image generation content policies. Actively replace terms that could be flagged as sensitive, explicit, or violent with safe, descriptive alternatives that preserve the visual intent. For example, describe 'blood' as red paint splashes or viscous red liquid, and rephrase potentially sensitive materials like 'rubber' to a neutral term such as elastic polymer.
\end{lstlisting}

\subsubsection{Video Analysis Prompts} \textit{Character Presence.}
\begin{lstlisting}[label={lst:character_json}]
JSON Output Structure:
[
  {
    "timestamp_start": <number>,
    "timestamp_end": <number>,
    "surrounding_character": {
      "description": "[Describes a single instance of a recurring, non-protagonist character type. e.g., 'Stormtrooper in white armor', 'Shambling zombie']",
      "motion": "[moving|static]",
      "quantity": "[a_few|several|many|countless]",
      "size": "[Given that an ant's size is 0.02 and a normal human's size is 1, return the expected size.]"
    }
  }
]

Module-Specific Rules:
- Crucial Rule: This field MUST be `null` if the non-protagonist is a single, unique individual (like a main villain), a named character, or if no other characters are present. It must also be `null` if the character is highly similar to the `surrounding_object`.
- AI Policy Compliance: In `surrounding_character.description`, avoid depicting graphic or sensitive content. Replace it with a neutral alternative (e.g., "mutilated body part" becomes "twisted dark root").
- Completeness Rule: If `description` has a value, all other fields in `surrounding_character` must also have values.
\end{lstlisting}

\subsubsection{Video Analysis Prompts} \textit{Body Transformation.}
\begin{lstlisting}[label={lst:body_json}]
JSON Output Structure:
[
  {
    "timestamp_start": <number>,
    "timestamp_end": <number>,
    "user_character": "[Describe the hand or arm of the central character. Focus on gloves, armor, tattoos, or unique skin. e.g., 'black, armored tactical glove', 'Blue-skinned Na'vi hand', 'covered in mud']"
  }
]

Module-Specific Rules:
- Focus: Describe the overall 'skin' or covering of the hand/arm. You MUST ignore and exclude small accessories like rings, bracelets, or watches.
- Absence of Features: If no hand/arm is visible, or if it is a plain, bare human hand/arm, state `null`.
\end{lstlisting}

\subsubsection{Video Analysis Prompts} \textit{Lighting Effects (Overall Lighting and Color).}
\begin{lstlisting}[label={lst:lnc_json}]
JSON Output Structure:
[
  {
    "timestamp_start": <number>,
    "timestamp_end": <number>,
    "lighting_and_color": {
      "brightness_level": "[very_dark|dark|normal|bright|overexposed]",
      "color_mood": "[very_cool|cool|neutral|warm|very_warm]",
      "dominant_color_hex": "#RRGGBB"
    }
  }
]
\end{lstlisting}

\subsubsection{Video Analysis Prompts} \textit{Lighting Effects (Effective Light).}
\begin{lstlisting}[label={lst:elight_json}]
JSON Output Structure:
[
  {
    "timestamp_start": <number>,
    "timestamp_end": <number>,
    "effective_light": {
      "type": "[blinking|search_light|explosion]",
      "color_hex": "#RRGGBB",
      "motion": "[static|sweeping|random]",
      "intensity": "[high|medium|low]"
    }
  }
]

Module-Specific Rules:
- If there are no identifiable, specific, special lighting effects within the segment, the value for the `effective_light` field should be `null`.
\end{lstlisting}

\subsubsection{Augmentation Generation Prompts} \textit{Image Generation.}
\begin{lstlisting}[label={lst:surface_json}]
Make an image that will be used as a {surface type} texture in unity. Seamless, tileable pattern texture, photorealistic, extremely detailed, flat and uniform, evenly lit, no depth, no object shape, no shadows, no highlights, diffuse lighting only, texture-only, 2D
\end{lstlisting}

\subsubsection{Augmentation Generation Prompts} \textit{Object Generation.}
\begin{lstlisting}[label={lst:3dprompt_json}]
Describe the following {object | character} in a detailed prompt to generate an image well. Description SHOULD BE START with "A full-body, photorealistic, highly detailed depiction of a {object | character}, isolated against a plain white background.", Do not include any additional objects in the description. The image should include the whole posture of the {object | character}, and the background should be white, the length of description should be less than 900, \n {object | character} : {keywords} in the movie {movie name}'
\end{lstlisting}

\end{document}